\documentclass[%
 reprint,
 amsmath,amssymb,
 aps,
 superscriptaddress,
]{revtex4-1}

\usepackage{graphicx}
\usepackage{dcolumn}
\usepackage{bm}
\usepackage{placeins}
\usepackage[export]{adjustbox}
\usepackage{ amssymb }
\usepackage{rotating}
\usepackage{lineno}
\usepackage{tabularx}
\usepackage[utf8]{inputenc}
\usepackage[T1]{fontenc}

\begin{document}


\title{Constraining the $\bar{p}/p$ Ratio in TeV Cosmic Rays with Observations of the Moon Shadow by HAWC}
\author{A.U.~Abeysekara}\affiliation{Department of Physics and Astronomy, University of Utah, Salt Lake City, UT, USA }
\author{A.~Albert}\affiliation{Physics Division, Los Alamos National Laboratory, Los Alamos, NM, USA }
\author{R.~Alfaro}\affiliation{Instituto de F\'{i}sica, Universidad Nacional Autónoma de México, Ciudad de Mexico, Mexico }
\author{C.~Alvarez}\affiliation{Universidad Autónoma de Chiapas, Tuxtla Gutiérrez, Chiapas, México}
\author{R.~Arceo}\affiliation{Universidad Autónoma de Chiapas, Tuxtla Gutiérrez, Chiapas, México}
\author{J.C.~Arteaga-Velázquez}\affiliation{Universidad Michoacana de San Nicolás de Hidalgo, Morelia, Mexico }
\author{D.~Avila Rojas}\affiliation{Instituto de F\'{i}sica, Universidad Nacional Autónoma de México, Ciudad de Mexico, Mexico }
\author{H.A.~Ayala Solares}\affiliation{Department of Physics, Pennsylvania State University, University Park, PA, USA }
\author{E.~Belmont-Moreno}\affiliation{Instituto de F\'{i}sica, Universidad Nacional Autónoma de México, Ciudad de Mexico, Mexico }
\author{S.Y.~BenZvi}\affiliation{Department of Physics \& Astronomy, University of Rochester, Rochester, NY , USA }
% ...


\collaboration{HAWC Collaboration}

\date{\today}

\begin{abstract}
An indirect measurement of the antiproton flux in cosmic rays is possible as the particles undergo deflection by the geomagnetic field. This effect can be measured by studying the deficit in the flux, or shadow, created by the Moon as it absorbs cosmic rays that are headed towards the Earth. The shadow is  displaced from the actual position of the Moon due to geomagnetic deflection, which is a function of the energy and charge of the cosmic rays. The displacement provides a natural tool for momentum/charge discrimination that can be used to study the composition of cosmic rays. Using 33 months of data comprising more than 80 billion cosmic rays measured by the High Altitude Water Cherenkov (HAWC) observatory, we have analyzed the Moon shadow to search for TeV antiprotons in cosmic rays. We present our first upper limits on the $\bar{p}/p$ fraction, which in the absence of any direct measurements, provide the tightest available constraints of $\sim1\%$ on the antiproton fraction for energies between 1 and 10 TeV.
\end{abstract}

\pacs{Valid PACS appear here}
\maketitle

\section{\label{sec:level1}Introduction}

Precision measurements of the cosmic-ray spectrum have brought an increased focus on antiparticles as a valuable tool for studying fundamental physics at very high energies. In standard models of cosmic-ray propagation, antiparticles such as $e^+$, $\bar{p}$ and $^3\overline{\text{He}}$ are produced as secondary species when primary cosmic-ray protons collide with interstellar gas in the Galaxy \cite{1475-7516-2015-12-039}. This picture is consistent with measurements of the antiproton to proton ratio between 10 GeV and 60 GeV made by several experiments, including BESS \cite{2002cosp...34E1239M}, HEAT \cite{PhysRevLett.87.271101}, CAPRICE \cite{Weber:1997zwa}, PAMELA \cite{2009PhRvL.102e1101A} and AMS-02 \cite{PhysRevLett.117.091103}.  At higher energies the ratio of secondary to primary components is an important testing ground for hitherto undiscovered sources of cosmic rays.

Measurements of the fluxes of individual species up to a few hundred GeV have revealed spectral features that are at odds with the predictions of propagation models \cite{PhysRevLett.117.091103,1742-6596-718-5-052012,1992ApJ...394..174G} that take into account diffusion, energy losses and gains, and particle production and disintegration   \cite{Moskalenko:2003kq,1988A&A...202....1S}. While adequately explaining some secondary to primary ratios such as B/C, the models do not seem to produce enough antiparticles to match the observations of recent high-statistics experiments \cite{2014PhRvD..89d3013C,2002ApJ...565..280M,1742-6596-384-1-012016,2017arXiv170906507B,2017ApJ...844L..26T,2016arXiv161204001L}. The latest data from the Alpha Magnetic Spectrometer (AMS-02) shows that the antiproton to proton ratio $\bar{p}/p$ is independent of rigidity (momentum divided by electric charge) between 10 GV and 450 GV  \cite{PhysRevLett.117.091103}, whereas in pure secondary production the ratio is expected to decrease with increasing rigidity \cite{2002ApJ...565..280M}. The antiproton to positron flux ratio $\bar{p}/e^+$ is also constant above 30 GV \cite{PhysRevLett.117.091103}, which is inconsistent with the different energy loss rates suffered by $\bar{p}$ and $e^+$ in the interstellar medium. \\
 
The observed excesses in fluxes of antiparticles could be due to unaccounted for astrophysical sources,decay or annihilation of exotic particles in physics beyond the standard model, or simply a reflection of uncertainties in our knowledge of the interstellar medium (ISM) and the interaction cross-sections used in secondary production models \cite{2017arXiv170906507B,PhysRevLett.117.091103}.

The search for new sources of antiparticles also makes antiprotons a potential target for indirect detection of dark matter. Annihilating or decaying dark matter may produce abundant $\bar{p}$ in hadronization processes that show up as an excess above the secondary-particle background in the spectrum \cite{Cirelli:2013hv}. A sharp cut-off in the fraction of antiprotons at a given energy could signal a dark matter particle in the same mass range undergoing annihilation \cite{Fornengo:2013xda}. The flattening of the $\bar{p}/p$ ratio at a few hundred GeV has led to lower limits on dark matter mass $m_{\chi}$ up to $\sim2$ TeV \cite{PhysRevD.92.055027,2015arXiv150407230L}, leaving the multi-TeV range open as testing ground for different scenarios.

Supernova remnants (SNRs) can also contribute to the $\bar{p}$ flux, resulting in a smooth increase in $\bar{p}/p$ until a maximum cut-off energy where it flattens \cite{PhysRevLett.103.081103,2041-8205-791-2-L22}. Depending on the age of the supernova remnant, the maximum acceleration energy can be $\mathcal{O}$(10 TeV), making TeV antiprotons important probes of astrophysical sources \cite{PhysRevLett.103.081103,PhysRevD.95.063021}. Characterizing the secondary antiparticle spectra across all accessible energies is therefore a well-motivated problem.

Measuring cosmic-ray antiprotons is a challenge owing to their very low flux at high energies and the difficulty of charge separation amongst hadrons in cosmic-ray detectors. While balloon and satellite experiments have good charge-sign resolution, they are limited in their exposure and their maximum energy sensitivity \cite{2007NIMPA.579.1034A, AGUILAR2002331}. AMS-02, for example, has provided direct measurements of the antiproton fraction up to 450 GeV. Ground based air shower arrays, with their large effective areas, can probe higher energies but are limited in their capability to identify individual primary particles. One approach to circumvent this problem is to study the deficit produced by the Moon in the cosmic-ray flux. The observation of this deficit or the Moon shadow is a common technique used by ground based cosmic-ray detectors to calibrate their angular resolution and pointing accuracy \cite{2014ApJ...796..108A, 2013arXiv1305.6811I}. Moreover, the position of the shadow is offset from the true location of the Moon due to the deflection of cosmic rays in the geomagnetic field. As a result, observations of the shadow can be used for momentum and charge-based separation of cosmic rays \cite{URBAN1990223,2007APh....28..137T,2012PhRvD..85b2002B}. By observing the Moon shadow, the ARGO-YBJ, MILAGRO, Tibet AS-$\gamma$ and the L3 collaborations estimated upper limits on the $\bar{p}/p$ ratio above 1 TeV at a few percent level \cite{2012PhRvD..85b2002B, 2011PhDT........70C, 2007APh....28..137T, Achard:2005az}.

The High Altitude Water Cherenkov Observatory (HAWC) is one of the very few operational ground-based experiments that can extend the $\bar{p}/p$ limits further into the very high energy regime. In this paper we use the measured Moon shadow to obtain the most constraining upper limits on the $\bar{p}/p$ ratio at energies between 1 TeV and 10 TeV. The paper is structured as follows: Section II describes the HAWC detector and the procedure of data selection. Section III discusses how the Moon shadow can be used to separate antiprotons from protons and also how to infer the experimental sensitivity of this measurement. Section IV shows the results of the search and the $95\%$ upper limits on the $\bar{p}/p$ ratio. Systematic uncertainties are also discussed in Section IV.  Section V provides an outlook and concludes the paper.

%
%

\section{\label{sec:level1}High Altitude Water Cherenkov Observatory}
\subsection{\label{sec:level2}The Detector}
The HAWC Observatory, located at an altitude of 4100 m above sea-level at Sierra Negra, Mexico, is a wide field-of-view detector array for TeV gamma rays and cosmic rays. It consists of 300 water Cherenkov detectors (WCDs) laid out over 22,000 m$^2$. Each WCD is a tank 7.3 m in diameter and 4.5 m in height filled with 180,000 liters of purified water. Four  upward-facing large-area photomultiplier tubes (PMTs) are anchored to the bottom of each WCD. Extensive air showers produced by incoming cosmic rays and gamma rays can trigger the PMTs as the cascade of secondary particles passes through the WCDs. \\

HAWC's nominal trigger rate is 25kHz, with the vast majority of triggers being due to air shower events. For this analysis, we use a multiplicity condition of at least 75 channels(PMTs) to be hit within a 150 ns time window to sort events as candidate air showers. After determining the effective charge in each PMT \cite{2017ApJ...843...39A}, any incorrectly calibrated triggers are removed and the events are reconstructed to obtain a lateral fit to the distribution of charge on the array. Combining this with the hit times of the PMTs allows us to infer shower parameters such as direction, location of the shower core, energy of the primary particle and particle type (cosmic ray or gamma ray). 

For estimating the energy of a cosmic-ray shower, we search a set of probability tables containing the lateral distribution of hits for a range of simulated proton energies and zenith angles. A likelihood value for each PMT is extracted from the table for a given shower with reconstructed zenith angle and core position. For each simulated bin of energy, the likelihood values are summed for all PMTs. The best estimate of energy corresponds to the bin with the maximum likelihood \cite{2017arXiv171000890H}. A complete description of the hardware and the data reconstruction methods used can be found in \cite{Abeysekara:2013qka,2017ApJ...843...39A}.

\subsection{\label{sec:level3}Simulations}
The event reconstruction and detector calibration make use of simulated extensive air showers generated with the CORSIKA package (v 7.40) \cite{1998cmcc.book.....H}, using the QGSJet-II-03 \cite{Ostapchenko:2004ss} model for hadronic interactions. This is followed by a GEANT4 (v4.10) simulation of secondary particles interacting with the HAWC array \cite{AGOSTINELLI2003250}. Custom software is then used to model the detector response taking into account the PMT efficiencies and noise in the readout channels \cite{Abeysekara:2013qka, 2017ApJ...843...39A}.

The cosmic-ray spectrum used in the simulations includes eight primary species (H, He, C, O, Ne, Mg, Si, Fe) with abundances based on the measurements by satellite and balloon experiments like CREAM \cite{2011ApJ...728..122Y}, PAMELA \cite{2011Sci...332...69A}, ATIC-2 \cite{2009BRASP..73..564P} and AMS \cite{PhysRevLett.114.171103}. The fluxes are parameterized using broken power law fits \cite{2017arXiv171000890H}. In addition, we also calculate the geomagnetic deflection of each cosmic-ray species. This is done by backtracing particles in the geomagnetic field from the location of HAWC to the Moon. The magnetic field is described by the most recent International Geomagnetic Reference Field (IGRF) model \cite{Thebault2015}. The detailed implementation of the particle propagation is described in \cite{2017arXiv171000890H}. By comparing the observed deflection with the expected results from the simulation, we can validate the energy scale and the pointing accuracy of the detector. 

\subsection{\label{sec:level3}The Data}
This work uses data collected by HAWC between November 2014 and August 2017. To ensure optimal reconstruction and energy estimation, only the events with a zenith angle of less than $45^\circ$ are used. Additional cuts reject events with shower cores far from the array \cite{2017arXiv171000890H,2017ApJ...843...39A}. The data are divided into energy bins from 1 TeV to 100 TeV with a width of $0.2$ in $\log_{10}{(E/\text{GeV)}}$. Over 81 billion cosmic rays survive these stringent quality cuts as shown in table \ref{tab:data}.

\begin{table}[]
\centering
\begin{tabular}{c|c|c}
\textbf{Bin} & \textbf{log($\mathbf{E}$/GeV)} & \textbf{Events/ $\mathbf{10^9}$} \\
\hline
0            & 3.0 - 3.2                      & 3.49                   \\
1            & 3.2 - 3.4                      & 17.67                 \\
2            & 3.4 - 3.6                      & 18.98                  \\
3            & 3.6 - 3.8                      & 13.50                  \\
4            & 3.8 - 4.0                      & 11.21                  \\
5            & 4.0 - 4.2                     & 7.63                   \\
6            & 4.2 - 4.4                    & 4.45                   \\
7            & $>$ 4.4                    & 4.44                   \\
\end{tabular}
\caption{Reconstructed energy and number of events in each bin after applying the data quality cuts.}
\label{tab:data}
\end{table}

\section{\label{sec:level1}Moon Shadow and the Search for Antiprotons}
\subsection{\label{sec:level2}Observation of the Moon by HAWC}
\begin{figure*}[htb!]
\centering
  \makebox[\textwidth][c]{\begin{tabular}{@{}ccc@{}}
    \includegraphics[width=.35\textwidth]{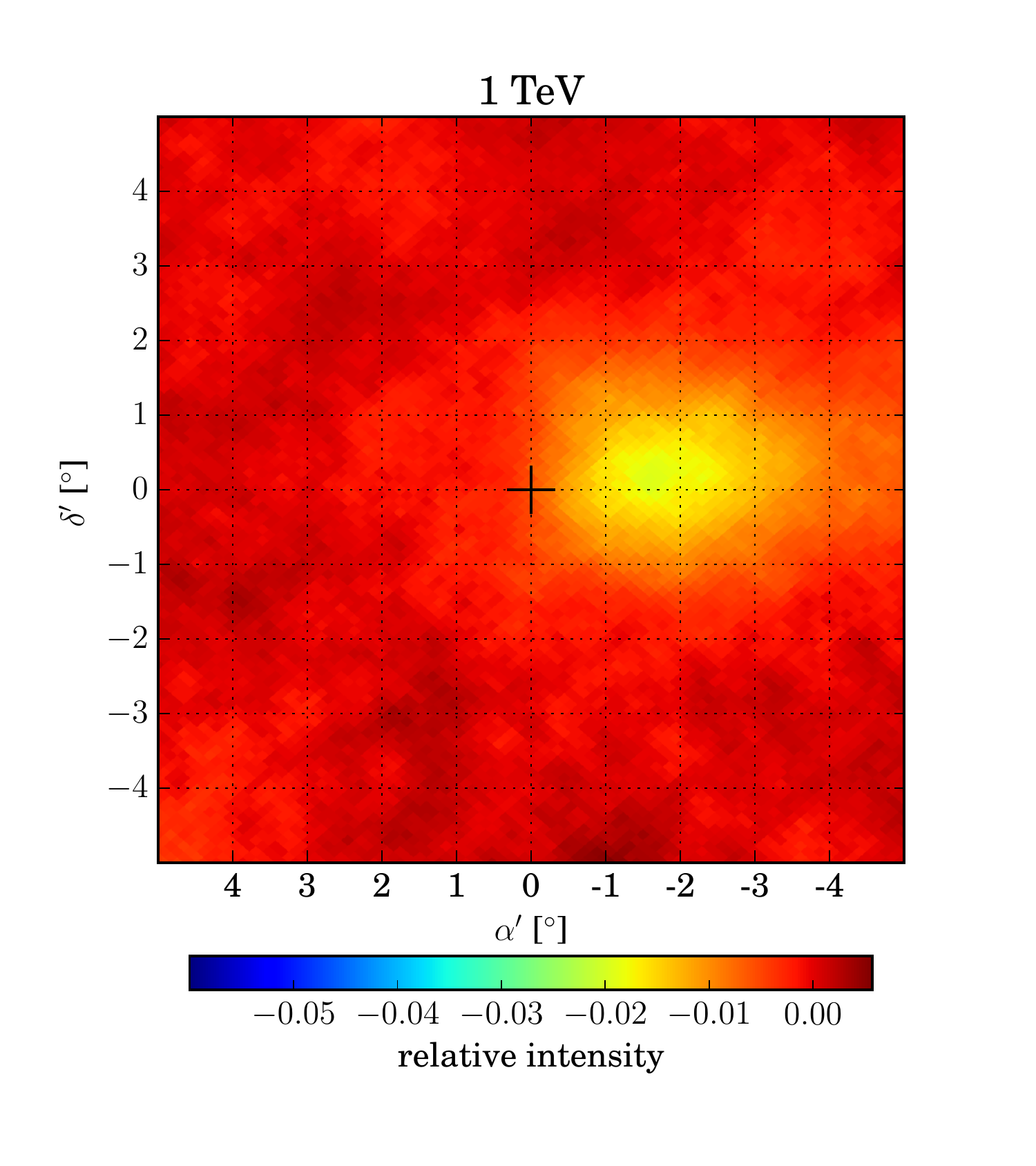} &
    \includegraphics[width=.35\textwidth]{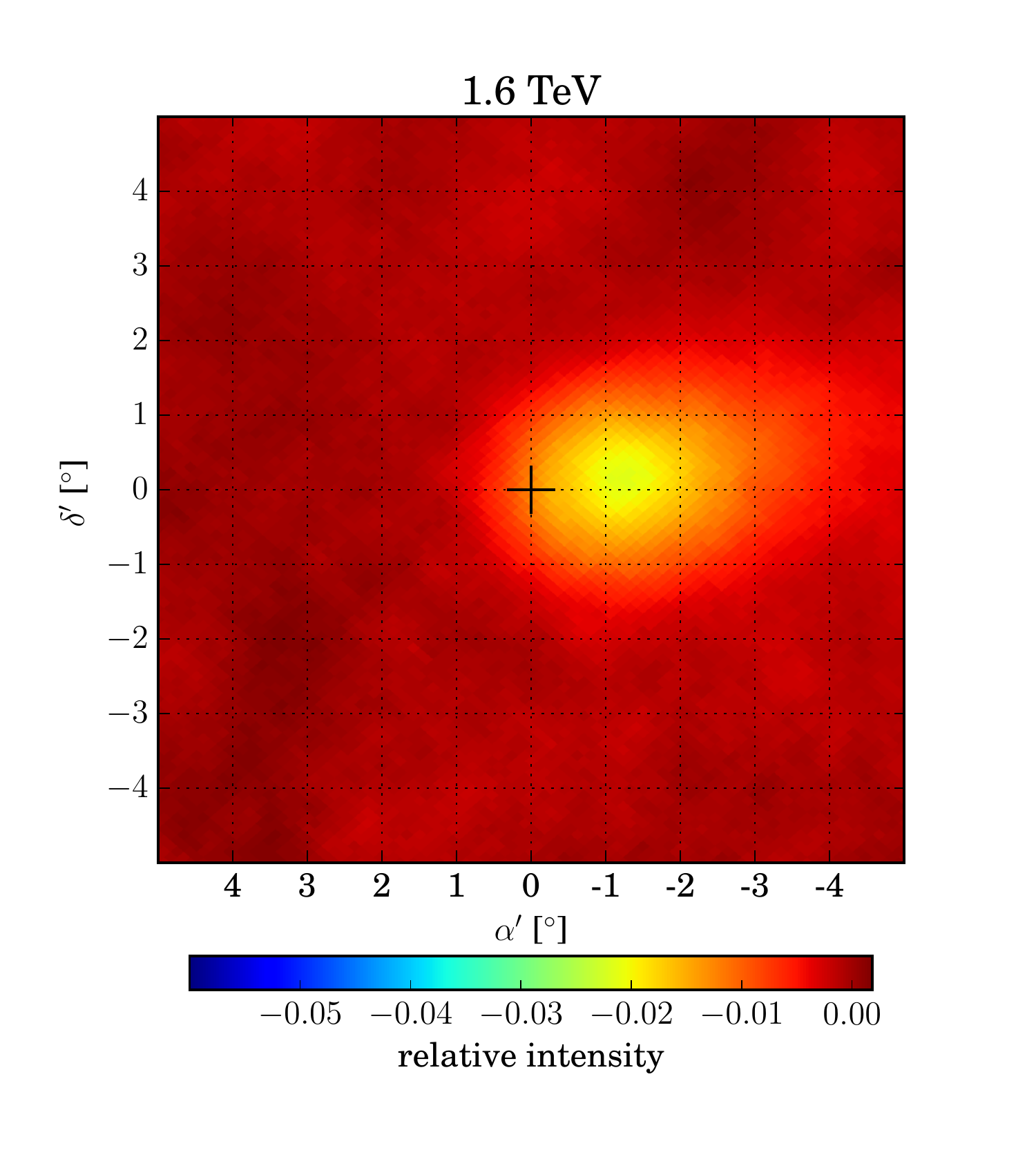} &
    \includegraphics[width=.35\textwidth]{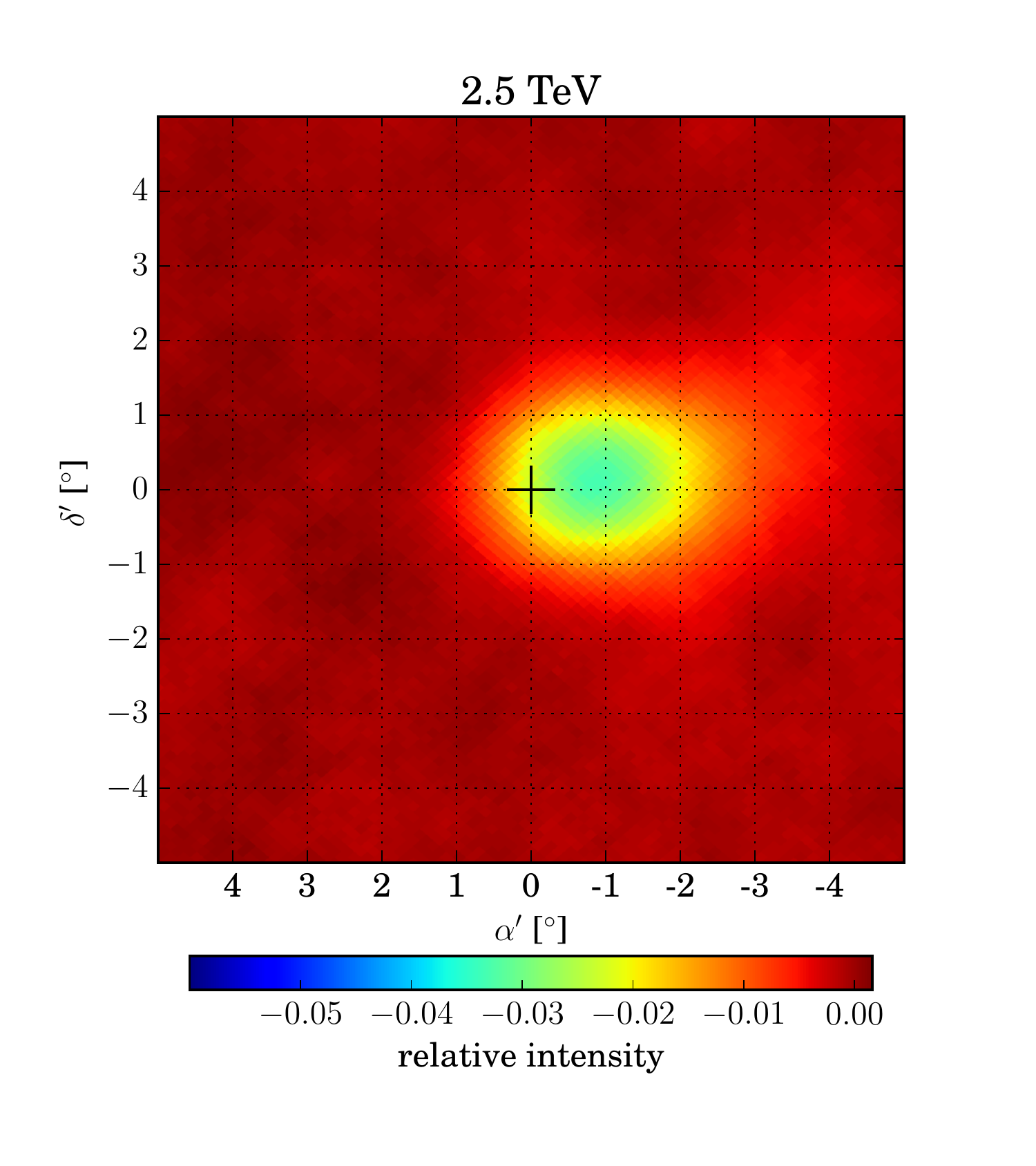} \\
    \includegraphics[width=.35\textwidth]{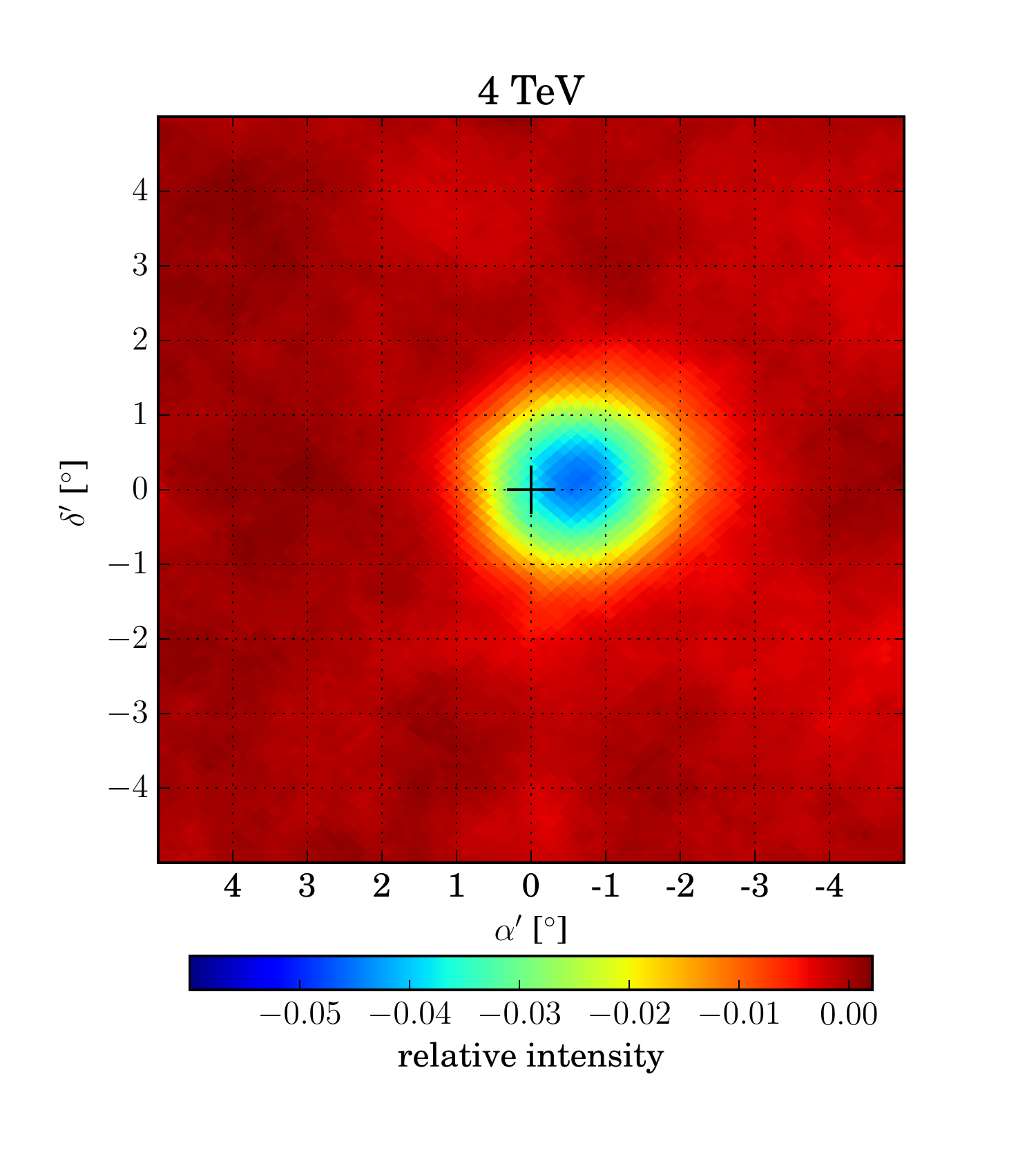} &
    \includegraphics[width=.35\textwidth]{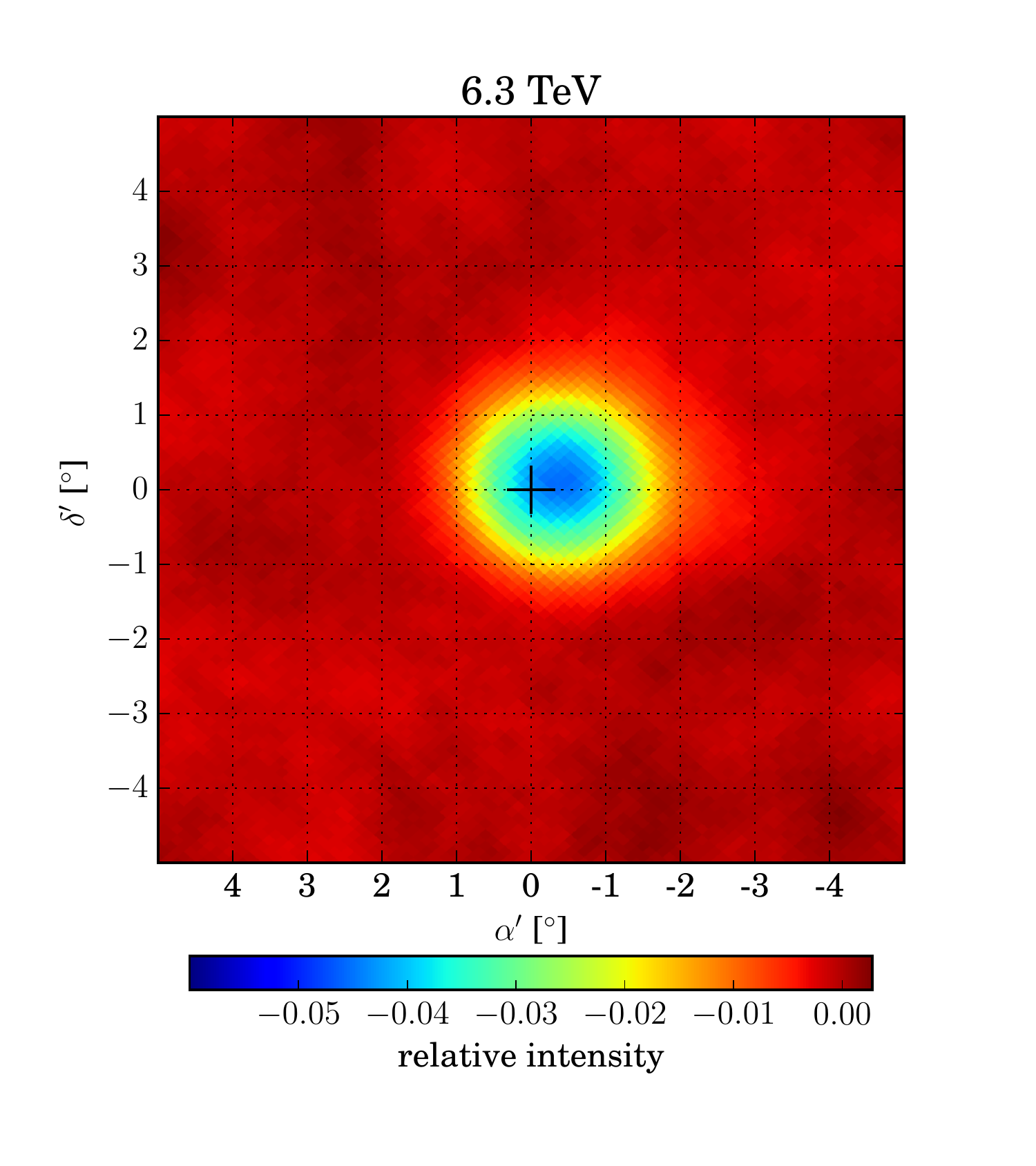} &
    \includegraphics[width=.35\textwidth]{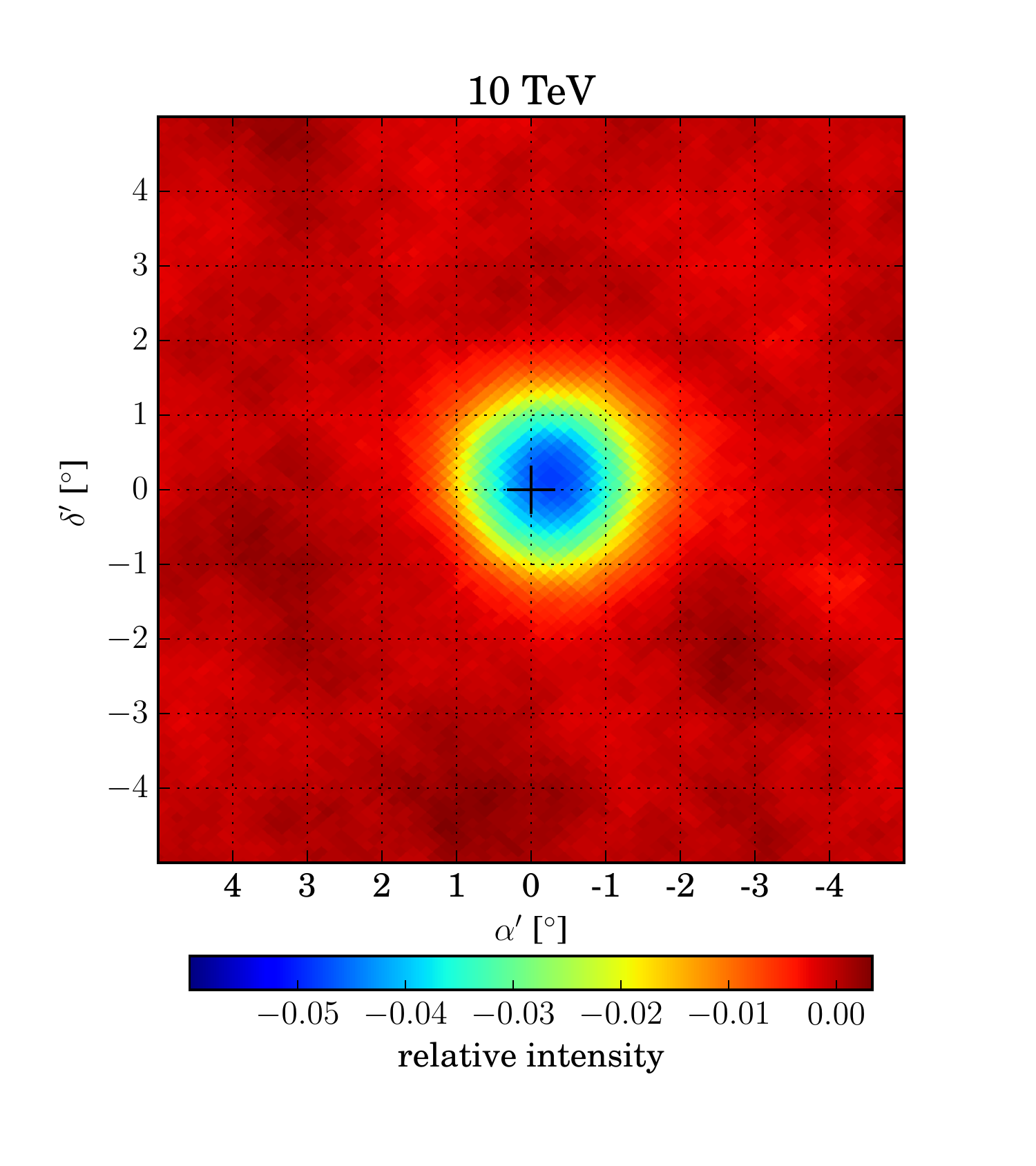} \\
    \includegraphics[width=.35\textwidth]{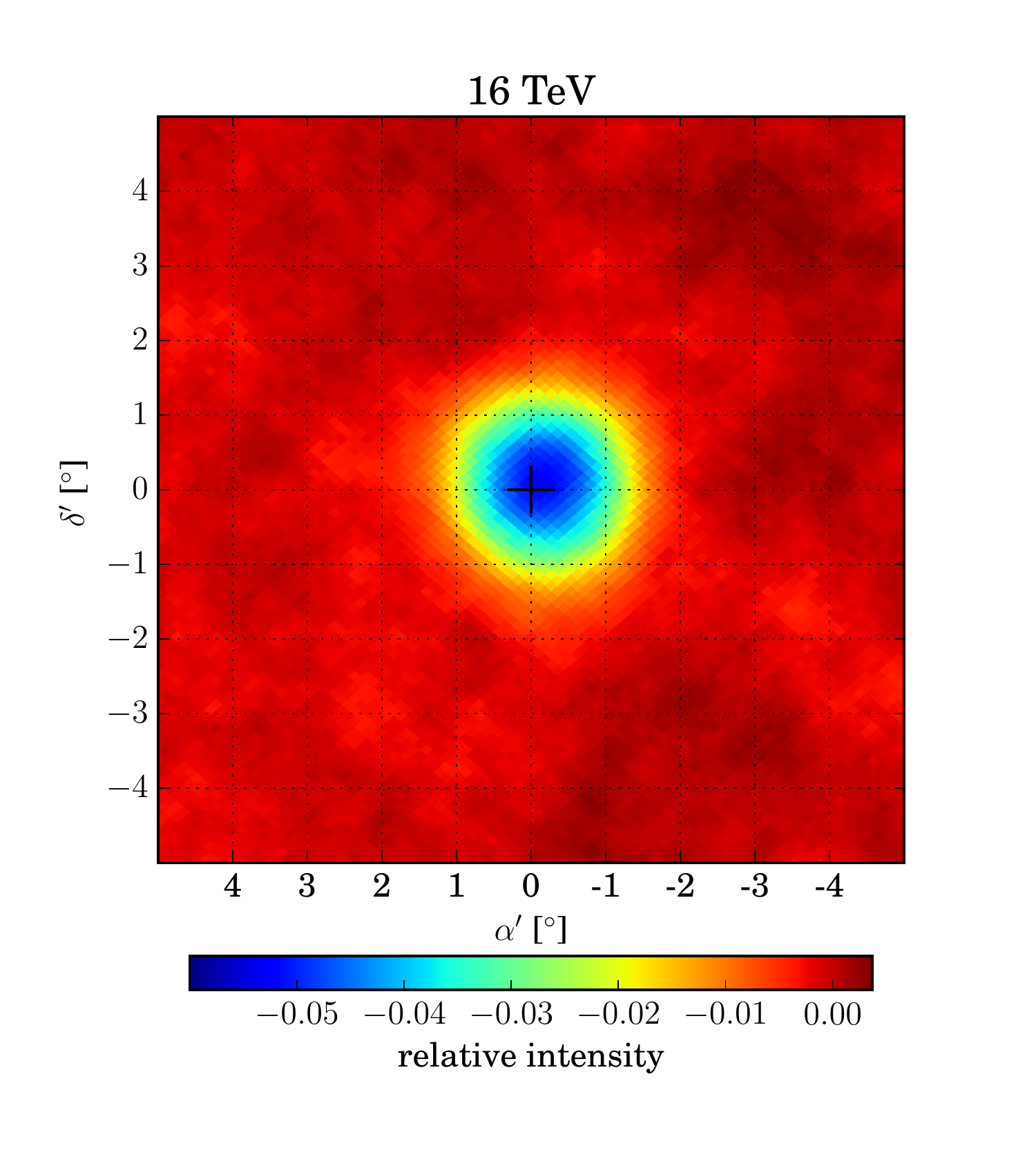} &
    \includegraphics[width=.35\textwidth]{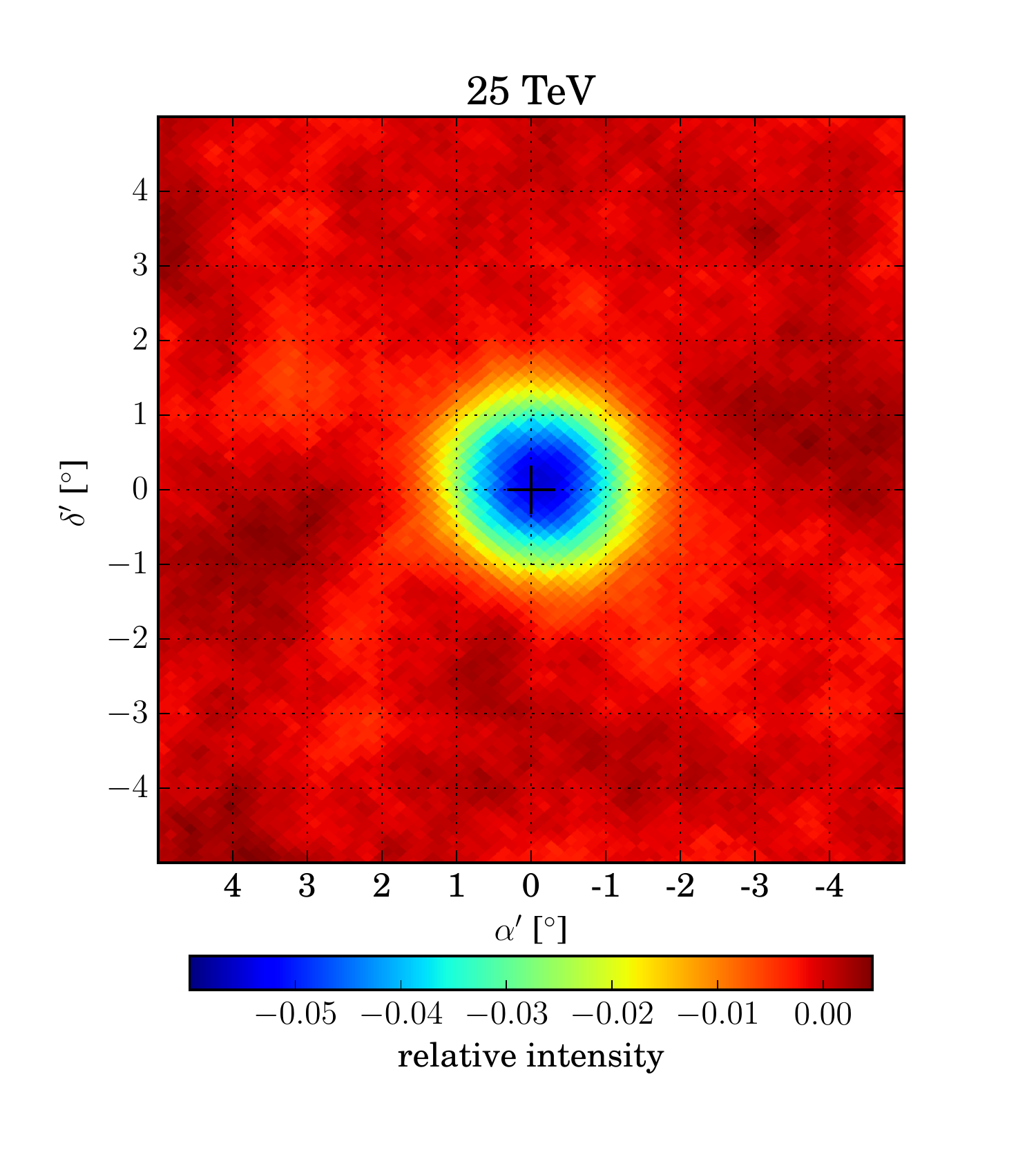} 
  \end{tabular}}
  \caption{The cosmic-ray Moon shadow at different energies in 33 months of data from HAWC. The maps have been smoothed by a $1^\circ$ top-hat function to visually enhance the shadow. The black cross indicates the actual position of the Moon in the moon-centered coordinates. The displacement in the centroid of the shadow due to geomagnetic deflection is highest at 1 TeV, $1.9^\circ$ in R.A and $0.3^\circ$ in declination. The offset in both directions decreases with energy approaching $(0.21 \pm 0.01)^\circ$ in RA and $(0.05 \pm 0.02)^\circ$ in Dec at 10 TeV.}
 \label{fig:all shadow bins}
\end{figure*}

We analyze the cosmic-ray flux by producing a sky-map of the data. The sky is divided into a grid of pixels of equal area in equatorial coordinates using the HEALPix library \cite{2005ApJ...622..759G}. Each pixel is centered at a right ascension and declination given by $(\alpha,\delta)$ and covers an angular width of about $0.1^\circ$. The map-making procedure quantifies the excess or deficit of cosmic-ray counts in every pixel with respect to an isotropic background. We define the relative intensity $\delta I$ as the fractional excess or deficit of counts in each pixel, 
\begin{equation}
\label{eq:ri}
\delta I = \frac{N(\alpha_i,\delta_i) - \langle N(\alpha_i,\delta_i)\rangle}{\langle N(\alpha_i,\delta_i)\rangle}
\end{equation}
where $N(\alpha_i,\delta_i)$ is the number of events in the data map and $\langle N(\alpha_i,\delta_i)\rangle$ is the counts in the isotropic reference map: the background distribution calculated using the method of Direct Integration \cite{2003ApJ...595..803A}. A significance $\sigma$ is also assigned to each pixel. The significance is a measure of the deviation of the data in each bin from the expectation of the isotropic map, and is calculated according to the techniques in Li\&Ma 1983 \cite{1983ApJ...272..317L}. 
To focus on the Moon, we subtract the calculated equatorial coordinates of the Moon from the coordinates of each event so that the final map is centered on the equatorial position of the Moon $(\alpha' = \alpha - \alpha_{\text{moon}}, \delta' = \delta - \delta_{\text{moon}})$ - the Moon being located at $(0,0)$ after the transformation. The Moon blocks the incoming cosmic rays, creating a deficit in the observed signal as shown in Fig. \ref{fig:all shadow bins}. This deficit or ``Moon shadow" is displaced from the Moon's actual position at $(\alpha'=0,\delta'=0)$ because of the deflection of cosmic rays in the earth's magnetic field. The expected angular deflection of a hadronic particle of charge $Z$ and energy $E$ at the location of HAWC is,
\begin{equation}
\label{eq4}
\delta \omega \simeq 1.6^{\circ}  Z (E/\text{1000 GeV})^{-1},
\end{equation}
which was obtained from simulations \cite{Abeysekara:2013qka,2017arXiv171000890H}.  We fit the shape of the shadow to an asymmetric 2D Gaussian as discussed in \ref{sec:level2} and use the centroid at $(\Delta\alpha',\Delta\delta')$ to describe the offset in position. The shape of the shadow is smeared along right ascension. This is because the reconstructed energy bins have a finite width, resulting in a broad distribution of geomagnetic deflections. The evolution of the shadow's width with energy is also a demonstration of the angular resolution of the detector - the angular width of the Moon-disc being $0.5^\circ$. 

\FloatBarrier
 
\begin{figure}[h!]
\centering
\includegraphics[width=0.52\textwidth]{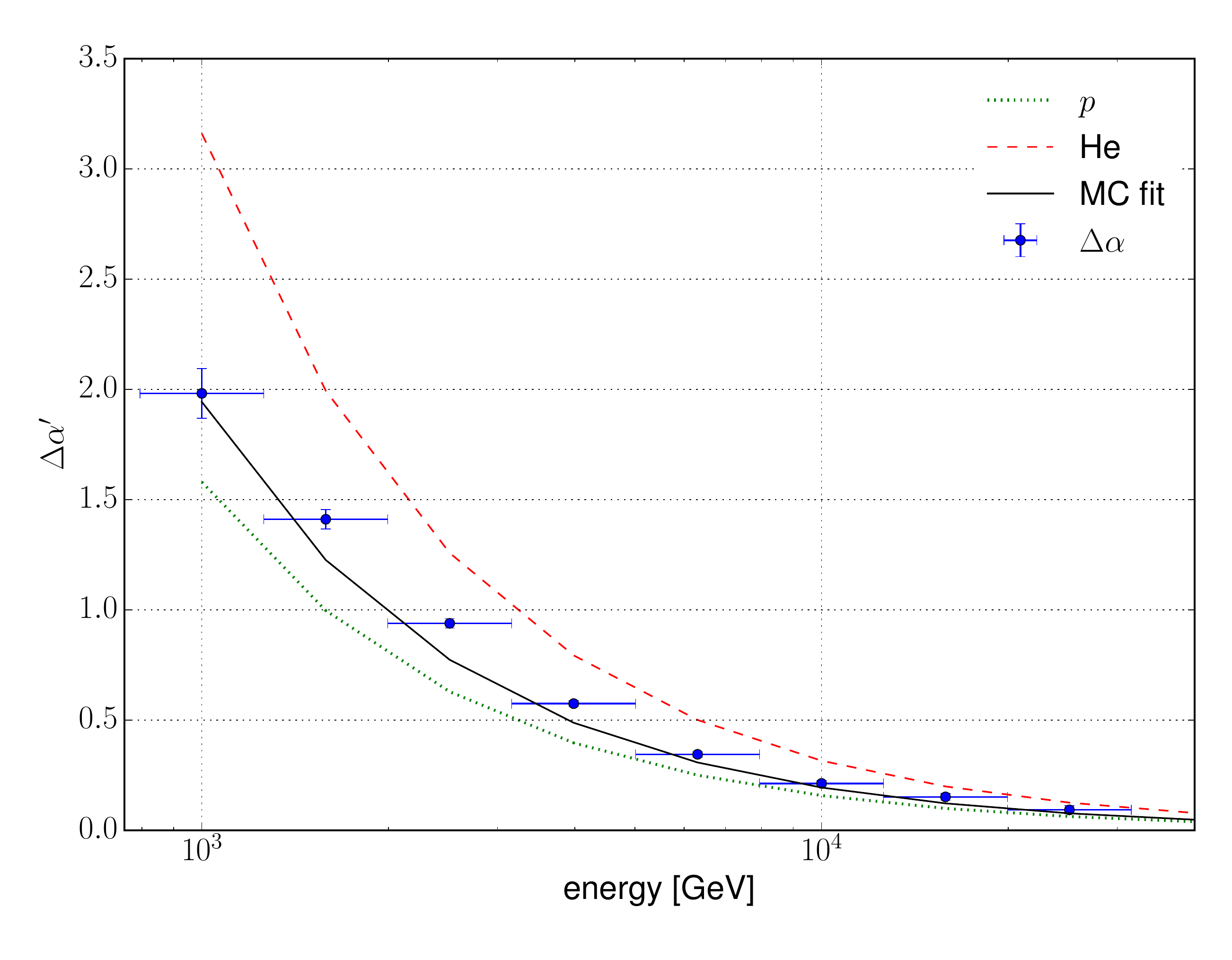}
\caption{The deflection of Moon shadow in right ascension as a function of energy in 33 months of data from HAWC. The dotted and dashed lines show the estimated deflection for pure protons and helium nuclei spectra respectively. The solid line is a fit to the mean deflection obtained from simulation. The blue points show the observed HAWC data.}
\label{fig:deltaRA}
\end{figure}

Fig. \ref{fig:deltaRA} illustrates the fit offset in right ascension as a function of energy. The expected offset of $p$ and He nuclei is also shown. We fit the same function from Eq.~\eqref{eq4} to the observed data and calculate $Z = 1.30 \pm 0.02$, obtaining an approximate estimate of the composition of the spectrum. Assuming $Z$ is an average of $p$ and He charges weighted by their abundance in the data, with negligible contribution from heavier elements, we estimate that about $(70 \pm 2)\%$ of the measured primary cosmic ray flux below 10 TeV is protons. While this fraction is only a rough estimate of the relative abundance of protons to helium, it is consistent with our detector efficiency for the assumed composition models which are based on direct measurements above 100 GeV \cite{2017arXiv171000890H}.

\subsection{\label{sec:level2}Finding a $\bar{p}$ shadow}
\begin{figure}[h]

\centering
\includegraphics[width=0.52\textwidth]{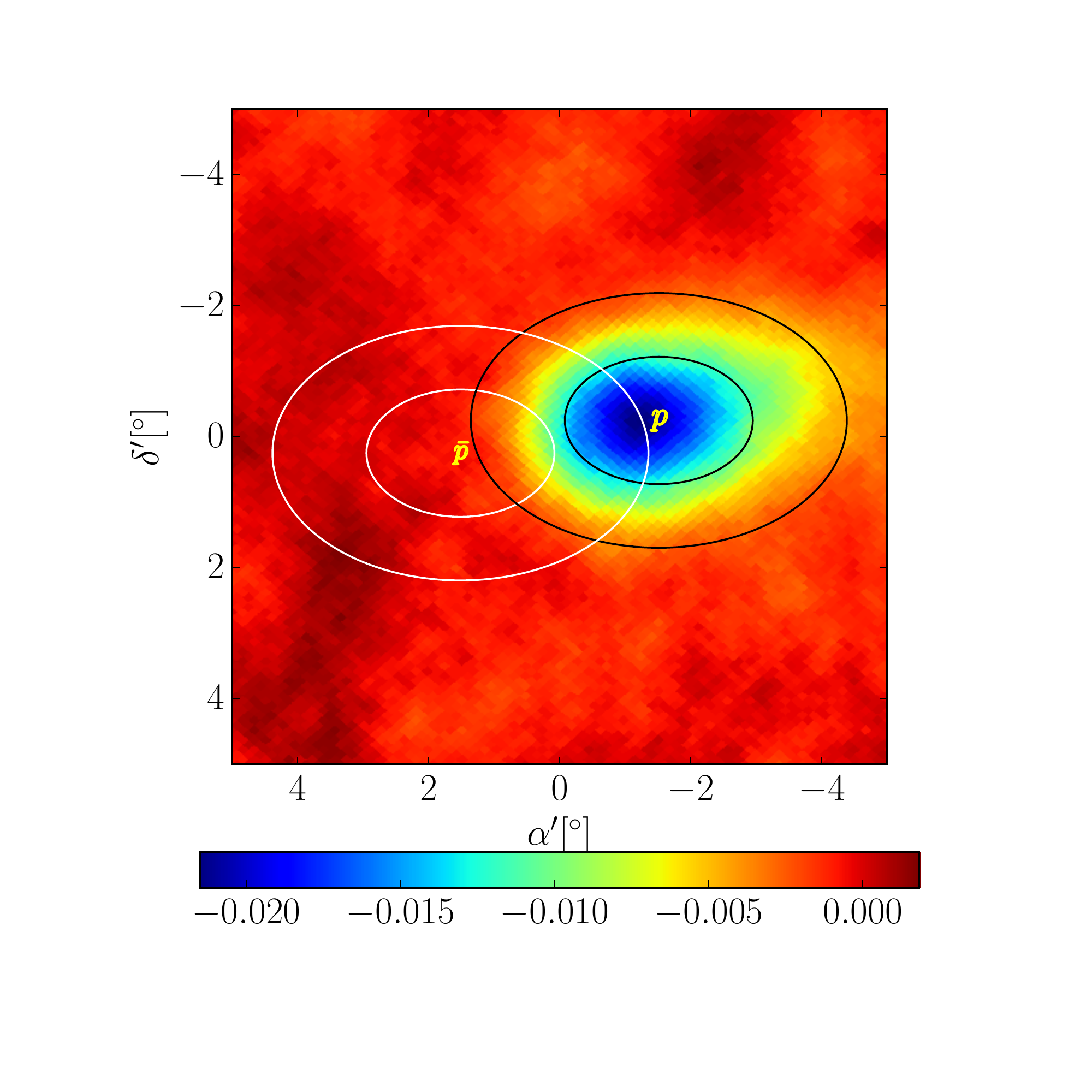}
\caption{The observed proton shadow at $1.6$ TeV, with 1$\sigma$ and 2$\sigma$ width contours of the fitted Gaussian overlaid. The white ellipses show the expected position of an antiproton shadow obtained by a $180^\circ$ rotation about the origin.}
\label{fig:pbarbin2}
\end{figure}

The observed deflection of the Moon shadow to the negative values of $\alpha'$ in Fig. \ref{fig:all shadow bins} is due to the positively charged protons and He nuclei in the cosmic-ray flux. In principle, negatively charged particles would be deflected in the opposite direction, creating another Moon shadow in the positive $\alpha',\delta'$ quadrant as shown in Fig. \ref{fig:pbarbin2}. Hence, one can  search for antiparticles in the cosmic-ray flux by looking for a second deficit. Below we describe our search for a second, spatially distinct shadow in the data whose ``depth" or relative intensity is proportional to the flux of antiprotons blocked by the Moon.

We start with a 2D Gaussian function, Eq.~\eqref{eq:gauss}, to describe the shape of the deficit in the Moon shadow. There are six free parameters in the fit: the centroids $x_0$ and $y_0$ (or $\Delta\alpha',\Delta\delta'$), the widths $\sigma_x$ and $\sigma_y$, the tilt angle $\theta$ the shadow makes with the $\alpha'$ axis, and the amplitude $A$. The value of the function at each $(\alpha',\delta')$ or $(x,y)$ corresponds to the relative intensity at the respective coordinate. 
\begin{multline}
f_i(x,y) = A \exp(-a(x-x_0)^2 +2b(x-x_0)(y-y_0)\\- c(y-y_0)^2)
\label{eq:gauss}
\end{multline}
with\\
%
\begin{align*}
  a &= \frac{\cos^2{\theta}}{2\sigma_{x}^2} + \frac{\sin^2{\theta}}{2\sigma_{y}^2},
  &
  b &= \frac{-\sin{2\theta}}{4\sigma_{x}^2} + \frac{\sin{2\theta}}{4\sigma_{y}^2},
  \\
  c &= \frac{\sin^2{\theta}}{2\sigma_{x}^2} + \frac{\cos^2{\theta}}{2\sigma_{y}^2}.
\end{align*}
%
Assuming the data contain both a $p$ and $\bar{p}$ shadow, we fit a sum of two elliptical Gaussian functions to the map:
\begin{equation}
\label{eq:gaussian}
\delta I(x,y) = F_p(x,y) + F_{\bar{p}}(x,y) = F_p(x,y) + r\cdot F_p(x,y),
\end{equation}
This can be used to measure the ratio $\bar{p}/p$ (denoted by $r$) or place upper limits if no second shadow is observed. Considering that antiproton and proton spectra have similar functional behavior at high energies \cite{PhysRevLett.117.091103}, we assume that an antiproton shadow should be a symmetric counterpart of the proton shadow, with the same magnitudes of all parameters except the amplitude $A$, and reflected about the declination and the right ascension axes. This means that in principle if we know the shape and position of the proton shadow from data or Monte Carlo, then we also have that information for the antiproton shadow. For conservative limits, we assume that the shadows are almost purely due to $p$ and $\bar{p}$ with similar spectra, and a negligible fraction of heavier nuclei. The systematic uncertainty introduced by this assumption is discussed in Section \ref{sec:level3}.

To simplify the problem, we perform the fit in two steps. First, we fit only the proton shadow to a single Gaussian and obtain the best fit values for the six free parameters. Then we fit the antiproton shadow by fixing its width and position using the values obtained in step 1. The amplitude of $\bar{p}$ can be written as $r\cdot A$ where $r = \bar{p}/p$ is the ratio of antiprotons to protons. We then use a simple maximum likelihood to obtain the value of $r$.

\subsubsection{Likelihood fit}
\begin{figure*}[ht!]
\centering
\includegraphics[width=0.95\textwidth]{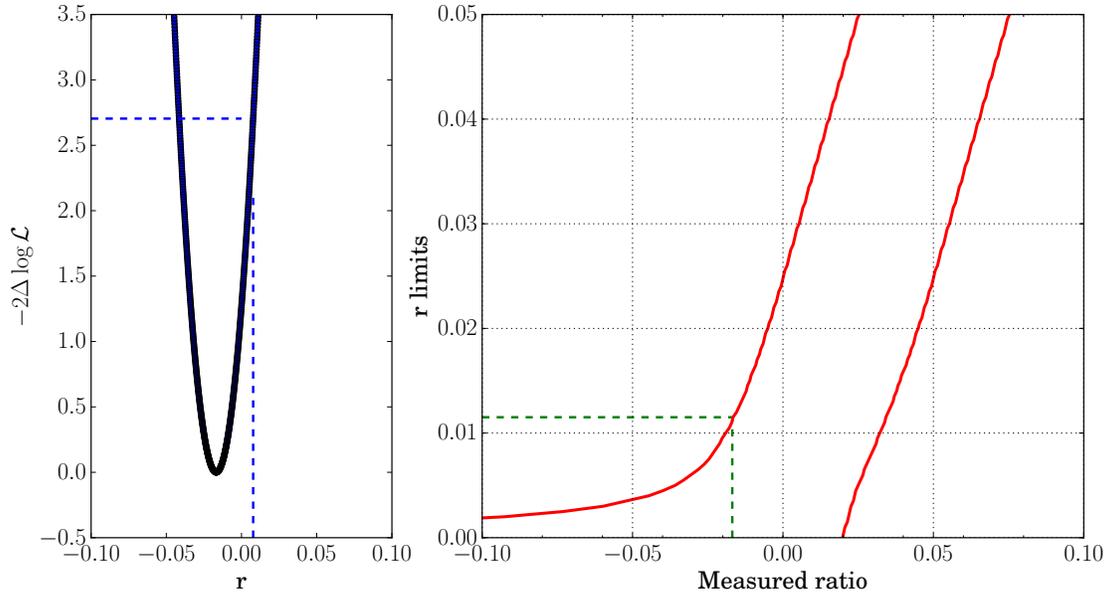}
\caption{\textit{Left}:The log-likelihood distribution for bin 2 (reconstructed median energy $= 2.5$ TeV). The blue dotted line shows the r interval corresponding to a $2\Delta \log \mathcal{L}$ of 2.71. \textit{Right}: The corresponding Feldman-Cousins interval \cite{1998PhRvD..57.3873F}. The green dashed line shows the Feldman-Cousins $95\%$ upper limit for the measurement shown on the left.}
\label{fig:LL}
\end{figure*}

To fit the antiproton shadow, we maximize the log-likelihood function

\begin{equation}
\label{eq1}
\log \mathcal{L} = -\frac{1}{2} \sum \limits_{i}^N \frac{(\delta I_i - \delta I(x_i,y_i,r))^2}{\sigma_i^2}
\end{equation}
where $\delta I_i$ is the relative intensity in the $i^{th}$ pixel of the Moon map for a given energy bin, $\sigma_i$ is the standard deviation in the relative intensity and $\delta I(x,y,r)$ is the superposition of two Gaussians as shown in Eq.~\eqref{eq:gaussian}. We minimize  $-\log \mathcal{L}$  using a grid search of over $10^4$  values of $r$. The resulting curve (Fig. \ref{fig:LL}) follows a Gaussian distribution. Its minimum corresponds to the optimal value $\hat{r}$. The contours of $\Delta \log \mathcal{L}(r)$ with respect to the minimum define the uncertainties in $\hat{r}$.\\

As illustrated for bin 2 in the left panel of Fig. \ref{fig:LL}, the Gaussian likelihood indicates a null result, i.e no antiproton shadow, with a negative value for $\hat{r}$ which is outside the physically allowed interval. To account for such underfluctuations of data and ensure that the reported $r$ at the $95\%$ confidence level (C.L.) is always positive, we calculate the upper limits following the Feldman \& Cousins  procedure \cite{1998PhRvD..57.3873F}. We use an implementation of the Feldman-Cousins confidence interval construction technique for a Gaussian that is truncated at zero, corresponding to our likelihood function for $r>0$, Fig. \ref{fig:LL}. The right panel in Fig. \ref{fig:LL} shows the $95\%$ upper and lower limits versus the measured values of $r$ in this scheme.\\

Applying the procedure described above to all bins with energy below 10 TeV, we obtain the upper limits shown in Fig. \ref{fig:limitsall}. We restrict the analysis to bin 5 (10 TeV) and below because the increased abundance of helium may bias the results at higher energies \cite{2011ApJ...728..122Y}. These issues will be addressed with improved particle discrimination techniques in future work.

\subsubsection{Sensitivity Calculation}
To study the effect of statistical fluctuations in the data on our computed upper limits, we calculate the sensitivity of HAWC to the antiproton shadow. In this context, the sensitivity refers to the average limit HAWC would obtain in an ensemble of similar experiments with no antiproton signal \cite{1998PhRvD..57.3873F}. This provides us with an independent range of \textit{minimum} values of $r$ that could be detected with at least a $95\%$ probability. In this way we can check for anomalous fluctuations in the background that may cause the measured upper limits to be significantly lower or greater than the sensitivity.\\

In this analysis, the absence of a shadow in any sampled region (other than the Moon) indicates that, barring fluctuations, the sampled data is consistent with the background. We compute the expected limit or sensitivity by searching for the antiproton shadow in 72 different regions --- each a circle of radius $5^\circ$ --- that are not within ten degrees of the Moon's position. We followed the procedure described in Section \ref{sec:level2} to fit the proton shadow at $\Delta \alpha',\Delta \delta'$. However, instead of $-\Delta \alpha,-\Delta \delta$ for the antiproton shadow, we used a random centroid at least $10^\circ$ away from the true Moon position. This ensures that we are only sampling off-source or background-only regions.  After repeating the fit on the 72 selected regions, we obtain a distribution of upper limits (yellow band in Fig. \ref{fig:limitsall}) or expected limits from only background. We notice that our 95\% upper limits fall within the range defined by the sensitivity of HAWC, alongside room for improvement with more statistics in future. 

\section{\label{sec:level1}Results}
\begin{figure*}[ht!]
\includegraphics[width=1.0\textwidth]{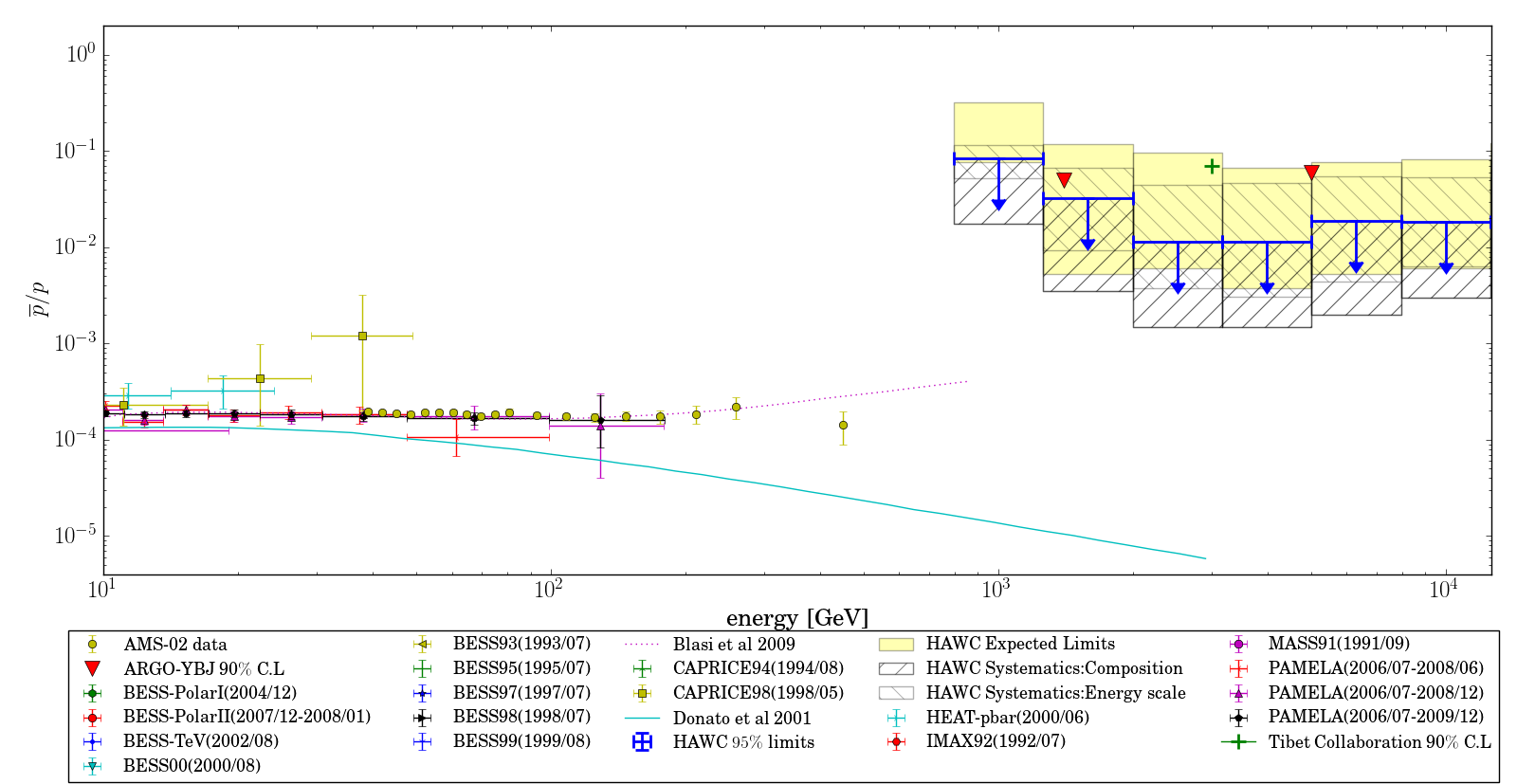}
\caption{Measurements of $\bar{p}/p$ in the GeV range and upper limits at the TeV scale. The yellow and shaded bands show HAWC sensitivity and systematic uncertainties respectively. The solid line shows the expected ratio from a purely secondary production of antiprotons \cite{PhysRevLett.102.071301}. The dotted line postulates primary antiproton production in supernovae \cite{PhysRevLett.103.081103}. Note that the other upper limits published above 1 TeV by ARGO-YBJ, L3 and Tibet AS-$\gamma$ are $90\%$ intervals while the HAWC limits are at the 95\% C.L.}
\label{fig:limitsall}
\end{figure*}

Table \ref{table:res2} lists the 95\%(90\%) upper limits from HAWC for different energy bins. With the high statistics available, the best results are $1.1\%$ at 95\% C.L. and $0.3\%$ at 90\% C.L. which is an order of magnitude improvement on previously published limits \cite{2012PhRvD..85b2002B,2007APh....28..137T, Achard:2005az}. Figure \ref{fig:limitsall} places our results in the context of past measurements and theoretical models. We are able to demonstrate HAWC's capability in performing an important constraining measurement at energies currently not accessible to direct detection experiments.

\begin{table}
\centering
\begin{tabular}{c|c|c|c}
$\mathbf{\log {(E/GeV)}}$ &  $\mathbf{\sigma_x}$ &  $\mathbf{\sigma_y}$ & $\mathbf{\bar{p}/p}$\textbf{[95(90) CL][\%]}\\
\hline
3.0                    					&$1.45\pm0.12$		&$0.90\pm0.07$			& 8.4 (6.6)               \\
3.2                      					&$1.24\pm0.05$		&$0.74\pm 0.02$		& 3.2 (2.5)               \\
3.4                      					&$0.93\pm0.02$		&$0.58\pm 0.01$		& 1.1 (0.3)             \\
3.6                       					&$0.65\pm0.01$		&$0.51\pm 0.01$		& 1.1 (0.8)                 \\
3.8                      		 			&$0.56\pm0.01$		&$0.49\pm 0.01$		& 1.9 (1.2)              \\
4.0                      		 			&$0.48\pm0.01$		&$0.47\pm 0.01$		& 1.9 (1.1)               \\

\end{tabular}
\caption{Estimated mean energies, shadow widths and HAWC 95\% and 90\% upper limits on the antiproton fraction.}
\label{table:res2}
\end{table}

\subsection{\label{sec:level3}Systematic Uncertainties}
\subsubsection{Composition}
One underlying assumption behind the fitting process is that the observed Moon shadow is predominantly due to incident cosmic-ray protons. However, the observed deflection and its comparison with simulations in Fig. \ref{fig:deltaRA} indicate a He component of up to $30\%$. This leads to an overestimate of the denominator in $\bar{p}/p$, making the limits on $r$ overly conservative. Several cosmic-ray experiments including AMS \cite{PhysRevLett.115.211101}, CREAM \cite{2011ApJ...728..122Y} and PAMELA \cite{2011Sci...332...69A} have measured a He fraction at 1 TeV around $25\%$ and show a hardening of the spectrum at multi-TeV energies \cite{2011ApJ...728..122Y}. The parameters of the fit showing the greatest sensitivity to He contamination are the amplitude and the offset in right ascension. We notice that the difference between the observed offset and pure proton offset is a small fraction of the width of the shadow at all energies explored in this study.\\

We investigated the systematic effect of varying the shadow parameters based on the proton spectrum of the composition model used in HAWC simulations. The upper limits were computed again after reducing the shadow deficit by $20\% - 30\%$ and shifting the offset to that expected from a pure proton shadow. These two factors were varied jointly, keeping all other parameters constant. Assuming no antihelium in the composition, we notice that a $20-25\%$ decrease in the shadow intensity along with a corresponding change in offset improves the limits by a factor of $2 - 8$ depending on energy. Figure \ref{fig:limitsall} shows the composition uncertainty in the shaded band, illustrating that our current results are conservative.
\subsubsection{Energy Reconstruction}The energy binning also contains systematic errors propagated from the probability tables used for estimating the energy of an air shower. The four dimensional tables have bins in zenith angle, the charge measured by a PMT, the distance of the PMT from the shower core and primary energy. The finite resolution and limited statistics in the tables contribute to the uncertainty in the likelihood and hence a bias between the true energy value and the reconstructed energy \cite{2017arXiv171000890H}. The trigger multiplicity and strict zenith angle cuts in this work were used to ensure the optimal performance of the energy estimator such that the bias in $\log_{10}{E}$ is restricted to the width of each energy bin \cite{2017PhDT........59H}. Any systematic shift in energy scale is directly propagated into the estimated flux \cite{2017arXiv171000890H} of protons. We studied this again by varying the shadow amplitude corresponding to the shift in flux that would result from a 10\% change in the energy scale. Figure \ref{fig:limitsall} shows that the corresponding shift in results falls within the range of expected limits.

The event reconstruction is also affected by shower fluctuations, the quantum efficiency and charge resolution of PMTs, and the interaction models used in array simulations \cite{2017arXiv171000890H,2017ApJ...843...39A}. In addition, the approximation of the Moon-disc with a 2D Gaussian may also produce a bias in the calculated deficit in different regions of the shadow. However, the systematic contribution of these effects on the estimated flux is of the order $5\%$ \cite{2017arXiv171000890H}, leaving the He contamination and energy scaling as the dominant sources of uncertainty.

 \section{\label{sec:level1}Conclusions}Probing the antiproton spectrum at TeV energies is an important prelude to developing a consistent theory to explain the production and propagation of secondary cosmic rays. The HAWC Observatory, with its continuous operation and sensitivity to TeV cosmic rays can constrain the $\bar{p}$ fraction. We achieve this by using the high-significance observation of the Moon shadow offset in position as a template for an antiproton shadow. The shape of the shadow is described by a two dimensional Gaussian with the ratio of $\bar{p}/p$ as a key parameter of the fit. With no observed antiproton shadow, we are able to place upper limits on $\bar{p}/p$ up to 10 TeV. The limits of 1.1\% at 2.5 TeV and 4 TeV, and $1.9\%$ at 10 TeV set an experimental bound that any models predicting a rise in the $\bar{p}/p$ fraction must satisfy \cite{PhysRevD.95.063021}. While these constraints are the strongest available at multi-TeV energies, we expect they can be improved with more HAWC data in the future, and can shed light on the secondary cosmic-ray background and potential signatures of new physics.

\bibliography{bib}{}

\begin{thebibliography}{49}%
\makeatletter
\providecommand \@ifxundefined [1]{%
 \@ifx{#1\undefined}
}%
\providecommand \@ifnum [1]{%
 \ifnum #1\expandafter \@firstoftwo
 \else \expandafter \@secondoftwo
 \fi
}%
\providecommand \@ifx [1]{%
 \ifx #1\expandafter \@firstoftwo
 \else \expandafter \@secondoftwo
 \fi
}%
\providecommand \natexlab [1]{#1}%
\providecommand \enquote  [1]{``#1''}%
\providecommand \bibnamefont  [1]{#1}%
\providecommand \bibfnamefont [1]{#1}%
\providecommand \citenamefont [1]{#1}%
\providecommand \href@noop [0]{\@secondoftwo}%
\providecommand \href [0]{\begingroup \@sanitize@url \@href}%
\providecommand \@href[1]{\@@startlink{#1}\@@href}%
\providecommand \@@href[1]{\endgroup#1\@@endlink}%
\providecommand \@sanitize@url [0]{\catcode `\\12\catcode `\$12\catcode
  `\&12\catcode `\#12\catcode `\^12\catcode `\_12\catcode `\%12\relax}%
\providecommand \@@startlink[1]{}%
\providecommand \@@endlink[0]{}%
\providecommand \url  [0]{\begingroup\@sanitize@url \@url }%
\providecommand \@url [1]{\endgroup\@href {#1}{\urlprefix }}%
\providecommand \urlprefix  [0]{URL }%
\providecommand \Eprint [0]{\href }%
\providecommand \doibase [0]{http://dx.doi.org/}%
\providecommand \selectlanguage [0]{\@gobble}%
\providecommand \bibinfo  [0]{\@secondoftwo}%
\providecommand \bibfield  [0]{\@secondoftwo}%
\providecommand \translation [1]{[#1]}%
\providecommand \BibitemOpen [0]{}%
\providecommand \bibitemStop [0]{}%
\providecommand \bibitemNoStop [0]{.\EOS\space}%
\providecommand \EOS [0]{\spacefactor3000\relax}%
\providecommand \BibitemShut  [1]{\csname bibitem#1\endcsname}%
\let\auto@bib@innerbib\@empty
\bibitem [{\citenamefont {Evoli}\ \emph {et~al.}(2015)\citenamefont {Evoli},
  \citenamefont {Gaggero},\ and\ \citenamefont
  {Grasso}}]{1475-7516-2015-12-039}%
  \BibitemOpen
  \bibfield  {author} {\bibinfo {author} {\bibfnamefont {C.}~\bibnamefont
  {Evoli}}, \bibinfo {author} {\bibfnamefont {D.}~\bibnamefont {Gaggero}}, \
  and\ \bibinfo {author} {\bibfnamefont {D.}~\bibnamefont {Grasso}},\ }\href
  {http://stacks.iop.org/1475-7516/2015/i=12/a=039} {\bibfield  {journal}
  {\bibinfo  {journal} {J. Cosmol. Astropart. Phys.}\ }\textbf {\bibinfo
  {volume} {2015}},\ \bibinfo {pages} {039} (\bibinfo {year}
  {2015})}\BibitemShut {NoStop}%
\bibitem [{\citenamefont {{Mitchell}}(2002)}]{2002cosp...34E1239M}%
  \BibitemOpen
  \bibfield  {author} {\bibinfo {author} {\bibfnamefont {J.}~\bibnamefont
  {{Mitchell}}},\ }in\ \href@noop {} {\emph {\bibinfo {booktitle} {34th COSPAR
  Scientific Assembly}}},\ \bibinfo {series} {COSPAR Meeting}, Vol.~\bibinfo
  {volume} {34}\ (\bibinfo {year} {2002})\BibitemShut {NoStop}%
\bibitem [{\citenamefont {Beach}\ \emph {et~al.}(2001)\citenamefont {Beach}
  \emph {et~al.}}]{PhysRevLett.87.271101}%
  \BibitemOpen
  \bibfield  {author} {\bibinfo {author} {\bibfnamefont {A.~S.}\ \bibnamefont
  {Beach}} \emph {et~al.},\ }\href {\doibase 10.1103/PhysRevLett.87.271101}
  {\bibfield  {journal} {\bibinfo  {journal} {Phys. Rev. Lett.}\ }\textbf
  {\bibinfo {volume} {87}},\ \bibinfo {pages} {271101} (\bibinfo {year}
  {2001})}\BibitemShut {NoStop}%
\bibitem [{\citenamefont {Weber}(1997)}]{Weber:1997zwa}%
  \BibitemOpen
  \bibfield  {author} {\bibinfo {author} {\bibfnamefont {N.}~\bibnamefont
  {Weber}},\ }\emph {\bibinfo {title} {{A Measurement of the Antiproton and
  Proton Fluxes in Cosmic Rays using the CAPRICE Experiment}}},\ \href
  {http://pamela.roma2.infn.it/index.php?option=com_docman&task=doc_view&gid=106}
  {Ph.D. thesis},\ \bibinfo  {school} {TU Stockholm} (\bibinfo {year}
  {1997})\BibitemShut {NoStop}%
\bibitem [{\citenamefont {{Adriani}}\ \emph {et~al.}(2009)\citenamefont
  {{Adriani}}, \citenamefont {{Barbarino}}, \citenamefont {{Bazilevskaya}}
  \emph {et~al.}}]{2009PhRvL.102e1101A}%
  \BibitemOpen
  \bibfield  {author} {\bibinfo {author} {\bibfnamefont {O.}~\bibnamefont
  {{Adriani}}}, \bibinfo {author} {\bibfnamefont {G.~C.}\ \bibnamefont
  {{Barbarino}}}, \bibinfo {author} {\bibfnamefont {G.~A.}\ \bibnamefont
  {{Bazilevskaya}}},  \emph {et~al.},\ }\href {\doibase
  10.1103/PhysRevLett.102.051101} {\bibfield  {journal} {\bibinfo  {journal}
  {Phys. Rev. Lett.}\ }\textbf {\bibinfo {volume} {102}},\ \bibinfo {eid}
  {051101} (\bibinfo {year} {2009})},\ \Eprint {http://arxiv.org/abs/0810.4994}
  {arXiv:0810.4994} \BibitemShut {NoStop}%
\bibitem [{\citenamefont {{Aguilar}}\ \emph {et~al.}(2016)\citenamefont
  {{Aguilar}} \emph {et~al.}}]{PhysRevLett.117.091103}%
  \BibitemOpen
  \bibfield  {author} {\bibinfo {author} {\bibfnamefont {M.}~\bibnamefont
  {{Aguilar}}} \emph {et~al.} (\bibinfo {collaboration} {AMS Collaboration}),\
  }\href {\doibase 10.1103/PhysRevLett.117.091103} {\bibfield  {journal}
  {\bibinfo  {journal} {Phys. Rev. Lett.}\ }\textbf {\bibinfo {volume} {117}},\
  \bibinfo {pages} {091103} (\bibinfo {year} {2016})}\BibitemShut {NoStop}%
\bibitem [{\citenamefont {Fiandrini}(2016)}]{1742-6596-718-5-052012}%
  \BibitemOpen
  \bibfield  {author} {\bibinfo {author} {\bibfnamefont {E.}~\bibnamefont
  {Fiandrini}},\ }\href {http://stacks.iop.org/1742-6596/718/i=5/a=052012}
  {\bibfield  {journal} {\bibinfo  {journal} {Journal of Physics: Conference
  Series}\ }\textbf {\bibinfo {volume} {718}},\ \bibinfo {pages} {052012}
  (\bibinfo {year} {2016})}\BibitemShut {NoStop}%
\bibitem [{\citenamefont {{Gaisser}}\ and\ \citenamefont
  {{Schaefer}}(1992)}]{1992ApJ...394..174G}%
  \BibitemOpen
  \bibfield  {author} {\bibinfo {author} {\bibfnamefont {T.~K.}\ \bibnamefont
  {{Gaisser}}}\ and\ \bibinfo {author} {\bibfnamefont {R.~K.}\ \bibnamefont
  {{Schaefer}}},\ }\href {\doibase 10.1086/171568} {\bibfield  {journal}
  {\bibinfo  {journal} {Astrophys. J.}\ }\textbf {\bibinfo {volume} {394}},\
  \bibinfo {pages} {174} (\bibinfo {year} {1992})}\BibitemShut {NoStop}%
\bibitem [{\citenamefont {Moskalenko}\ \emph {et~al.}(2003)\citenamefont
  {Moskalenko}, \citenamefont {Strong}, \citenamefont {Mashnik},\ and\
  \citenamefont {Ormes}}]{Moskalenko:2003kq}%
  \BibitemOpen
  \bibfield  {author} {\bibinfo {author} {\bibfnamefont {I.~V.}\ \bibnamefont
  {Moskalenko}}, \bibinfo {author} {\bibfnamefont {A.~W.}\ \bibnamefont
  {Strong}}, \bibinfo {author} {\bibfnamefont {S.~G.}\ \bibnamefont {Mashnik}},
  \ and\ \bibinfo {author} {\bibfnamefont {J.~F.}\ \bibnamefont {Ormes}},\ }in\
  \href {http://www-rccn.icrr.u-tokyo.ac.jp/icrc2003/PROCEEDINGS/PDF/475.pdf}
  {\emph {\bibinfo {booktitle} {{Proceedings, 28th International Cosmic Ray
  Conference (ICRC 2003): Tsukuba, Japan, July 31-August 7, 2003}}}}\ (\bibinfo
  {year} {2003})\ p.\ \bibinfo {pages} {1921},\ \bibinfo {note}
  {[4,1921(2003)]},\ \Eprint {http://arxiv.org/abs/astro-ph/0306368}
  {arXiv:astro-ph/0306368 [astro-ph]} \BibitemShut {NoStop}%
\bibitem [{\citenamefont {{Stephens}}\ and\ \citenamefont
  {{Golden}}(1988)}]{1988A&A...202....1S}%
  \BibitemOpen
  \bibfield  {author} {\bibinfo {author} {\bibfnamefont {S.~A.}\ \bibnamefont
  {{Stephens}}}\ and\ \bibinfo {author} {\bibfnamefont {R.~L.}\ \bibnamefont
  {{Golden}}},\ }\href@noop {} {\bibfield  {journal} {\bibinfo  {journal}
  {Astronomy and Astrophysics}\ }\textbf {\bibinfo {volume} {202}},\ \bibinfo
  {pages} {1} (\bibinfo {year} {1988})}\BibitemShut {NoStop}%
\bibitem [{\citenamefont {{Cholis}}\ and\ \citenamefont
  {{Hooper}}(2014)}]{2014PhRvD..89d3013C}%
  \BibitemOpen
  \bibfield  {author} {\bibinfo {author} {\bibfnamefont {I.}~\bibnamefont
  {{Cholis}}}\ and\ \bibinfo {author} {\bibfnamefont {D.}~\bibnamefont
  {{Hooper}}},\ }\href {\doibase 10.1103/PhysRevD.89.043013} {\bibfield
  {journal} {\bibinfo  {journal} {\prd}\ }\textbf {\bibinfo {volume} {89}},\
  \bibinfo {eid} {043013} (\bibinfo {year} {2014})},\ \Eprint
  {http://arxiv.org/abs/1312.2952} {arXiv:1312.2952 [astro-ph.HE]} \BibitemShut
  {NoStop}%
\bibitem [{\citenamefont {{Moskalenko}}\ \emph {et~al.}(2002)\citenamefont
  {{Moskalenko}}, \citenamefont {{Strong}}, \citenamefont {{Ormes}},\ and\
  \citenamefont {{Potgieter}}}]{2002ApJ...565..280M}%
  \BibitemOpen
  \bibfield  {author} {\bibinfo {author} {\bibfnamefont {I.~V.}\ \bibnamefont
  {{Moskalenko}}}, \bibinfo {author} {\bibfnamefont {A.~W.}\ \bibnamefont
  {{Strong}}}, \bibinfo {author} {\bibfnamefont {J.~F.}\ \bibnamefont
  {{Ormes}}}, \ and\ \bibinfo {author} {\bibfnamefont {M.~S.}\ \bibnamefont
  {{Potgieter}}},\ }\href {\doibase 10.1086/324402} {\bibfield  {journal}
  {\bibinfo  {journal} {Astrophys. J.}\ }\textbf {\bibinfo {volume} {565}},\
  \bibinfo {pages} {280} (\bibinfo {year} {2002})},\ \Eprint
  {http://arxiv.org/abs/astro-ph/0106567} {astro-ph/0106567} \BibitemShut
  {NoStop}%
\bibitem [{\citenamefont {Auchettl}\ and\ \citenamefont
  {Balázs}(2012)}]{1742-6596-384-1-012016}%
  \BibitemOpen
  \bibfield  {author} {\bibinfo {author} {\bibfnamefont {K.}~\bibnamefont
  {Auchettl}}\ and\ \bibinfo {author} {\bibfnamefont {C.}~\bibnamefont
  {Balázs}},\ }\href {http://stacks.iop.org/1742-6596/384/i=1/a=012016}
  {\bibfield  {journal} {\bibinfo  {journal} {Journal of Physics: Conference
  Series}\ }\textbf {\bibinfo {volume} {384}},\ \bibinfo {pages} {012016}
  (\bibinfo {year} {2012})}\BibitemShut {NoStop}%
\bibitem [{\citenamefont {{Blum}}\ \emph {et~al.}(2017)\citenamefont {{Blum}},
  \citenamefont {{Sato}},\ and\ \citenamefont
  {{Waxman}}}]{2017arXiv170906507B}%
  \BibitemOpen
  \bibfield  {author} {\bibinfo {author} {\bibfnamefont {K.}~\bibnamefont
  {{Blum}}}, \bibinfo {author} {\bibfnamefont {R.}~\bibnamefont {{Sato}}}, \
  and\ \bibinfo {author} {\bibfnamefont {E.}~\bibnamefont {{Waxman}}},\
  }\href@noop {} {\bibfield  {journal} {\bibinfo  {journal} {ArXiv e-prints}\ }
  (\bibinfo {year} {2017})},\ \Eprint {http://arxiv.org/abs/1709.06507}
  {arXiv:1709.06507 [astro-ph.HE]} \BibitemShut {NoStop}%
\bibitem [{\citenamefont {{Tomassetti}}\ and\ \citenamefont
  {{Oliva}}(2017)}]{2017ApJ...844L..26T}%
  \BibitemOpen
  \bibfield  {author} {\bibinfo {author} {\bibfnamefont {N.}~\bibnamefont
  {{Tomassetti}}}\ and\ \bibinfo {author} {\bibfnamefont {A.}~\bibnamefont
  {{Oliva}}},\ }\href {\doibase 10.3847/2041-8213/aa80da} {\bibfield  {journal}
  {\bibinfo  {journal} {Astrophys. J.}\ }\textbf {\bibinfo {volume} {844}},\
  \bibinfo {eid} {L26} (\bibinfo {year} {2017})},\ \Eprint
  {http://arxiv.org/abs/1707.06915} {arXiv:1707.06915 [astro-ph.HE]}
  \BibitemShut {NoStop}%
\bibitem [{\citenamefont {{Lin}}\ \emph {et~al.}(2016)\citenamefont {{Lin}},
  \citenamefont {{Bi}}, \citenamefont {{Feng}}, \citenamefont {{Yin}},\ and\
  \citenamefont {{Yu}}}]{2016arXiv161204001L}%
  \BibitemOpen
  \bibfield  {author} {\bibinfo {author} {\bibfnamefont {S.-J.}\ \bibnamefont
  {{Lin}}}, \bibinfo {author} {\bibfnamefont {X.-J.}\ \bibnamefont {{Bi}}},
  \bibinfo {author} {\bibfnamefont {J.}~\bibnamefont {{Feng}}}, \bibinfo
  {author} {\bibfnamefont {P.-F.}\ \bibnamefont {{Yin}}}, \ and\ \bibinfo
  {author} {\bibfnamefont {Z.-H.}\ \bibnamefont {{Yu}}},\ }\href@noop {}
  {\bibfield  {journal} {\bibinfo  {journal} {ArXiv e-prints}\ } (\bibinfo
  {year} {2016})},\ \Eprint {http://arxiv.org/abs/1612.04001} {arXiv:1612.04001
  [astro-ph.HE]} \BibitemShut {NoStop}%
\bibitem [{\citenamefont {Cirelli}\ and\ \citenamefont
  {Giesen}(2013)}]{Cirelli:2013hv}%
  \BibitemOpen
  \bibfield  {author} {\bibinfo {author} {\bibfnamefont {M.}~\bibnamefont
  {Cirelli}}\ and\ \bibinfo {author} {\bibfnamefont {G.}~\bibnamefont
  {Giesen}},\ }\href {\doibase 10.1088/1475-7516/2013/04/015} {\bibfield
  {journal} {\bibinfo  {journal} {J. Cosmol. Astropart. Phys.}\ }\textbf
  {\bibinfo {volume} {1304}},\ \bibinfo {pages} {015} (\bibinfo {year}
  {2013})},\ \Eprint {http://arxiv.org/abs/1301.7079} {arXiv:1301.7079
  [hep-ph]} \BibitemShut {NoStop}%
\bibitem [{\citenamefont {Fornengo}\ \emph {et~al.}(2014)\citenamefont
  {Fornengo}, \citenamefont {Maccione},\ and\ \citenamefont
  {Vittino}}]{Fornengo:2013xda}%
  \BibitemOpen
  \bibfield  {author} {\bibinfo {author} {\bibfnamefont {N.}~\bibnamefont
  {Fornengo}}, \bibinfo {author} {\bibfnamefont {L.}~\bibnamefont {Maccione}},
  \ and\ \bibinfo {author} {\bibfnamefont {A.}~\bibnamefont {Vittino}},\ }\href
  {\doibase 10.1088/1475-7516/2014/04/003} {\bibfield  {journal} {\bibinfo
  {journal} {J. Cosmol. Astropart. Phys.}\ }\textbf {\bibinfo {volume}
  {1404}},\ \bibinfo {pages} {003} (\bibinfo {year} {2014})},\ \Eprint
  {http://arxiv.org/abs/1312.3579} {arXiv:1312.3579 [hep-ph]} \BibitemShut
  {NoStop}%
\bibitem [{\citenamefont {Jin}\ \emph {et~al.}(2015)\citenamefont {Jin},
  \citenamefont {Wu},\ and\ \citenamefont {Zhou}}]{PhysRevD.92.055027}%
  \BibitemOpen
  \bibfield  {author} {\bibinfo {author} {\bibfnamefont {H.-B.}\ \bibnamefont
  {Jin}}, \bibinfo {author} {\bibfnamefont {Y.-L.}\ \bibnamefont {Wu}}, \ and\
  \bibinfo {author} {\bibfnamefont {Y.-F.}\ \bibnamefont {Zhou}},\ }\href
  {\doibase 10.1103/PhysRevD.92.055027} {\bibfield  {journal} {\bibinfo
  {journal} {Phys. Rev. D}\ }\textbf {\bibinfo {volume} {92}},\ \bibinfo
  {pages} {055027} (\bibinfo {year} {2015})}\BibitemShut {NoStop}%
\bibitem [{\citenamefont {{Lin}}\ \emph {et~al.}(2015)\citenamefont {{Lin}},
  \citenamefont {{Bi}}, \citenamefont {{Yin}},\ and\ \citenamefont
  {{Yu}}}]{2015arXiv150407230L}%
  \BibitemOpen
  \bibfield  {author} {\bibinfo {author} {\bibfnamefont {S.-J.}\ \bibnamefont
  {{Lin}}}, \bibinfo {author} {\bibfnamefont {X.-J.}\ \bibnamefont {{Bi}}},
  \bibinfo {author} {\bibfnamefont {P.-F.}\ \bibnamefont {{Yin}}}, \ and\
  \bibinfo {author} {\bibfnamefont {Z.-H.}\ \bibnamefont {{Yu}}},\ }\href@noop
  {} {\bibfield  {journal} {\bibinfo  {journal} {ArXiv e-prints}\ } (\bibinfo
  {year} {2015})},\ \Eprint {http://arxiv.org/abs/1504.07230} {arXiv:1504.07230
  [hep-ph]} \BibitemShut {NoStop}%
\bibitem [{\citenamefont {Blasi}\ and\ \citenamefont
  {Serpico}(2009)}]{PhysRevLett.103.081103}%
  \BibitemOpen
  \bibfield  {author} {\bibinfo {author} {\bibfnamefont {P.}~\bibnamefont
  {Blasi}}\ and\ \bibinfo {author} {\bibfnamefont {P.~D.}\ \bibnamefont
  {Serpico}},\ }\href {\doibase 10.1103/PhysRevLett.103.081103} {\bibfield
  {journal} {\bibinfo  {journal} {Phys. Rev. Lett.}\ }\textbf {\bibinfo
  {volume} {103}},\ \bibinfo {pages} {081103} (\bibinfo {year}
  {2009})}\BibitemShut {NoStop}%
\bibitem [{\citenamefont {Berezhko}\ and\ \citenamefont
  {Ksenofontov}(2014)}]{2041-8205-791-2-L22}%
  \BibitemOpen
  \bibfield  {author} {\bibinfo {author} {\bibfnamefont {E.~G.}\ \bibnamefont
  {Berezhko}}\ and\ \bibinfo {author} {\bibfnamefont {L.~T.}\ \bibnamefont
  {Ksenofontov}},\ }\href {http://stacks.iop.org/2041-8205/791/i=2/a=L22}
  {\bibfield  {journal} {\bibinfo  {journal} {Astrophys. J. Lett.}\ }\textbf
  {\bibinfo {volume} {791}},\ \bibinfo {pages} {L22} (\bibinfo {year}
  {2014})}\BibitemShut {NoStop}%
\bibitem [{\citenamefont {Huang}\ \emph {et~al.}(2017)\citenamefont {Huang},
  \citenamefont {Wei}, \citenamefont {Wu}, \citenamefont {Zhang},\ and\
  \citenamefont {Zhou}}]{PhysRevD.95.063021}%
  \BibitemOpen
  \bibfield  {author} {\bibinfo {author} {\bibfnamefont {X.-J.}\ \bibnamefont
  {Huang}}, \bibinfo {author} {\bibfnamefont {C.-C.}\ \bibnamefont {Wei}},
  \bibinfo {author} {\bibfnamefont {Y.-L.}\ \bibnamefont {Wu}}, \bibinfo
  {author} {\bibfnamefont {W.-H.}\ \bibnamefont {Zhang}}, \ and\ \bibinfo
  {author} {\bibfnamefont {Y.-F.}\ \bibnamefont {Zhou}},\ }\href {\doibase
  10.1103/PhysRevD.95.063021} {\bibfield  {journal} {\bibinfo  {journal} {Phys.
  Rev. D}\ }\textbf {\bibinfo {volume} {95}},\ \bibinfo {pages} {063021}
  (\bibinfo {year} {2017})}\BibitemShut {NoStop}%
\bibitem [{\citenamefont {{Ahn}}\ \emph {et~al.}(2007)\citenamefont {{Ahn}},
  \citenamefont {{Allison}}, \citenamefont {{Bagliesi}}, \citenamefont
  {{Beatty}} \emph {et~al.}}]{2007NIMPA.579.1034A}%
  \BibitemOpen
  \bibfield  {author} {\bibinfo {author} {\bibfnamefont {H.~S.}\ \bibnamefont
  {{Ahn}}}, \bibinfo {author} {\bibfnamefont {P.}~\bibnamefont {{Allison}}},
  \bibinfo {author} {\bibfnamefont {M.~G.}\ \bibnamefont {{Bagliesi}}},
  \bibinfo {author} {\bibfnamefont {J.~J.}\ \bibnamefont {{Beatty}}},  \emph
  {et~al.},\ }\href {\doibase 10.1016/j.nima.2007.05.203} {\bibfield  {journal}
  {\bibinfo  {journal} {Nuclear Instruments and Methods in Physics Research A}\
  }\textbf {\bibinfo {volume} {579}},\ \bibinfo {pages} {1034} (\bibinfo {year}
  {2007})}\BibitemShut {NoStop}%
\bibitem [{\citenamefont {Aguilar}\ \emph {et~al.}(2002)\citenamefont {Aguilar}
  \emph {et~al.}}]{AGUILAR2002331}%
  \BibitemOpen
  \bibfield  {author} {\bibinfo {author} {\bibfnamefont {M.}~\bibnamefont
  {Aguilar}} \emph {et~al.},\ }\href {\doibase
  https://doi.org/10.1016/S0370-1573(02)00013-3} {\bibfield  {journal}
  {\bibinfo  {journal} {Physics Reports}\ }\textbf {\bibinfo {volume} {366}},\
  \bibinfo {pages} {331 } (\bibinfo {year} {2002})}\BibitemShut {NoStop}%
\bibitem [{\citenamefont {{Abeysekara}}\ \emph {et~al.}(2014)\citenamefont
  {{Abeysekara}} \emph {et~al.}}]{2014ApJ...796..108A}%
  \BibitemOpen
  \bibfield  {author} {\bibinfo {author} {\bibfnamefont {A.~U.}\ \bibnamefont
  {{Abeysekara}}} \emph {et~al.} (\bibinfo {collaboration} {HAWC
  Collaboration}),\ }\href {\doibase 10.1088/0004-637X/796/2/108} {\bibfield
  {journal} {\bibinfo  {journal} {Astrophys. J.}\ }\textbf {\bibinfo {volume}
  {796}},\ \bibinfo {eid} {108} (\bibinfo {year} {2014})},\ \Eprint
  {http://arxiv.org/abs/1408.4805} {arXiv:1408.4805 [astro-ph.HE]} \BibitemShut
  {NoStop}%
\bibitem [{\citenamefont {{Aartsen}}\ \emph {et~al.}(2013)\citenamefont
  {{Aartsen}}, \citenamefont {{Abbasi}} \emph {et~al.}}]{2013arXiv1305.6811I}%
  \BibitemOpen
  \bibfield  {author} {\bibinfo {author} {\bibfnamefont {M.~G.}\ \bibnamefont
  {{Aartsen}}}, \bibinfo {author} {\bibfnamefont {R.}~\bibnamefont {{Abbasi}}},
   \emph {et~al.} (\bibinfo {collaboration} {IceCube Collaboration}),\
  }\href@noop {} {\bibfield  {journal} {\bibinfo  {journal} {ArXiv e-prints}\ }
  (\bibinfo {year} {2013})},\ \Eprint {http://arxiv.org/abs/1305.6811}
  {arXiv:1305.6811 [astro-ph.HE]} \BibitemShut {NoStop}%
\bibitem [{\citenamefont {Urban}\ \emph {et~al.}(1990)\citenamefont {Urban},
  \citenamefont {Fleury}, \citenamefont {Lestienne},\ and\ \citenamefont
  {Plouin}}]{URBAN1990223}%
  \BibitemOpen
  \bibfield  {author} {\bibinfo {author} {\bibfnamefont {M.}~\bibnamefont
  {Urban}}, \bibinfo {author} {\bibfnamefont {P.}~\bibnamefont {Fleury}},
  \bibinfo {author} {\bibfnamefont {R.}~\bibnamefont {Lestienne}}, \ and\
  \bibinfo {author} {\bibfnamefont {F.}~\bibnamefont {Plouin}},\ }\href
  {\doibase https://doi.org/10.1016/0920-5632(90)90384-7} {\bibfield  {journal}
  {\bibinfo  {journal} {Nuclear Physics B - Proceedings Supplements}\ }\textbf
  {\bibinfo {volume} {14}},\ \bibinfo {pages} {223 } (\bibinfo {year}
  {1990})}\BibitemShut {NoStop}%
\bibitem [{\citenamefont {{Tibet As{$\gamma$}
  Collaboration}}(2007)}]{2007APh....28..137T}%
  \BibitemOpen
  \bibfield  {author} {\bibinfo {author} {\bibnamefont {{Tibet As{$\gamma$}
  Collaboration}}},\ }\href {\doibase 10.1016/j.astropartphys.2007.05.002}
  {\bibfield  {journal} {\bibinfo  {journal} {Astroparticle Physics}\ }\textbf
  {\bibinfo {volume} {28}},\ \bibinfo {pages} {137} (\bibinfo {year} {2007})},\
  \Eprint {http://arxiv.org/abs/0707.3326} {arXiv:0707.3326} \BibitemShut
  {NoStop}%
\bibitem [{\citenamefont {{Bartoli}}\ \emph {et~al.}(2012)\citenamefont
  {{Bartoli}} \emph {et~al.}}]{2012PhRvD..85b2002B}%
  \BibitemOpen
  \bibfield  {author} {\bibinfo {author} {\bibfnamefont {B.}~\bibnamefont
  {{Bartoli}}} \emph {et~al.},\ }\href {\doibase 10.1103/PhysRevD.85.022002}
  {\bibfield  {journal} {\bibinfo  {journal} {Phys. Rev. D}\ }\textbf {\bibinfo
  {volume} {85}},\ \bibinfo {eid} {022002} (\bibinfo {year} {2012})},\ \Eprint
  {http://arxiv.org/abs/1201.3848} {arXiv:1201.3848 [astro-ph.HE]} \BibitemShut
  {NoStop}%
\bibitem [{\citenamefont {{Christopher}}(2011)}]{2011PhDT........70C}%
  \BibitemOpen
  \bibfield  {author} {\bibinfo {author} {\bibfnamefont {G.~E.}\ \bibnamefont
  {{Christopher}}},\ }\emph {\bibinfo {title} {{Physics from the Very-High
  Energy Cosmic-Ray Shadows of the Moon and Sun with Milagro}}},\ \href@noop {}
  {Ph.D. thesis},\ \bibinfo  {school} {New York University} (\bibinfo {year}
  {2011})\BibitemShut {NoStop}%
\bibitem [{\citenamefont {{Abeysekara}}\ \emph {et~al.}(2017)\citenamefont
  {{Abeysekara}} \emph {et~al.}}]{2017ApJ...843...39A}%
  \BibitemOpen
  \bibfield  {author} {\bibinfo {author} {\bibfnamefont {A.~U.}\ \bibnamefont
  {{Abeysekara}}} \emph {et~al.} (\bibinfo {collaboration} {HAWC
  Collaboration}),\ }\href {\doibase 10.3847/1538-4357/aa7555} {\bibfield
  {journal} {\bibinfo  {journal} {Astrophys. J.}\ }\textbf {\bibinfo {volume}
  {843}},\ \bibinfo {eid} {39} (\bibinfo {year} {2017})},\ \Eprint
  {http://arxiv.org/abs/1701.01778} {arXiv:1701.01778 [astro-ph.HE]}
  \BibitemShut {NoStop}%
\bibitem [{\citenamefont {Alfaro}\ \emph {et~al.}(2017)\citenamefont {Alfaro}
  \emph {et~al.}}]{2017arXiv171000890H}%
  \BibitemOpen
  \bibfield  {author} {\bibinfo {author} {\bibfnamefont {R.}~\bibnamefont
  {Alfaro}} \emph {et~al.} (\bibinfo {collaboration} {HAWC Collaboration}),\
  }\href {\doibase 10.1103/PhysRevD.96.122001} {\bibfield  {journal} {\bibinfo
  {journal} {Phys. Rev. D}\ }\textbf {\bibinfo {volume} {96}},\ \bibinfo
  {pages} {122001} (\bibinfo {year} {2017})}\BibitemShut {NoStop}%
\bibitem [{\citenamefont {Abeysekara}\ \emph {et~al.}(2014)\citenamefont
  {Abeysekara} \emph {et~al.}}]{Abeysekara:2013qka}%
  \BibitemOpen
  \bibfield  {author} {\bibinfo {author} {\bibfnamefont {A.~U.}\ \bibnamefont
  {Abeysekara}} \emph {et~al.} (\bibinfo {collaboration} {HAWC
  Collaboration}),\ }\href@noop {} {\bibfield  {journal} {\bibinfo  {journal}
  {Braz. J. Phys.}\ }\textbf {\bibinfo {volume} {44}} (\bibinfo {year}
  {2014})},\ \Eprint {http://arxiv.org/abs/1310.0071} {arXiv:1310.0071
  [astro-ph.HE]} \BibitemShut {NoStop}%
\bibitem [{\citenamefont {{Heck}}\ \emph {et~al.}(1998)\citenamefont {{Heck}},
  \citenamefont {{Knapp}}, \citenamefont {{Capdevielle}}, \citenamefont
  {{Schatz}},\ and\ \citenamefont {{Thouw}}}]{1998cmcc.book.....H}%
  \BibitemOpen
  \bibfield  {author} {\bibinfo {author} {\bibfnamefont {D.}~\bibnamefont
  {{Heck}}}, \bibinfo {author} {\bibfnamefont {J.}~\bibnamefont {{Knapp}}},
  \bibinfo {author} {\bibfnamefont {J.~N.}\ \bibnamefont {{Capdevielle}}},
  \bibinfo {author} {\bibfnamefont {G.}~\bibnamefont {{Schatz}}}, \ and\
  \bibinfo {author} {\bibfnamefont {T.}~\bibnamefont {{Thouw}}},\ }\href@noop
  {} {\emph {\bibinfo {title} {CORSIKA: a Monte Carlo code to simulate
  extensive air showers., by Heck, D.; Knapp, J.; Capdevielle, J.~N.; Schatz,
  G.; Thouw, T..~ Forschungszentrum Karlsruhe GmbH, Karlsruhe (Germany)., Feb
  1998, V + 90 p., TIB Hannover, D-30167 Hannover (Germany).}}}\ (\bibinfo
  {year} {1998})\BibitemShut {NoStop}%
\bibitem [{\citenamefont {Ostapchenko}(2006)}]{Ostapchenko:2004ss}%
  \BibitemOpen
  \bibfield  {author} {\bibinfo {author} {\bibfnamefont {S.}~\bibnamefont
  {Ostapchenko}},\ }\bibfield  {booktitle} {\emph {\bibinfo {booktitle}
  {{Proceedings, 13th International Symposium on Very High-Energy Cosmic Ray
  Interactions (ISVHECRI 2004): Pylos, Greece, September 6-12, 2004}}},\ }\href
  {\doibase 10.1016/j.nuclphysbps.2005.07.026} {\bibfield  {journal} {\bibinfo
  {journal} {Nucl. Phys. Proc. Suppl.}\ }\textbf {\bibinfo {volume} {151}},\
  \bibinfo {pages} {143} (\bibinfo {year} {2006})},\ \Eprint
  {http://arxiv.org/abs/hep-ph/0412332} {arXiv:hep-ph/0412332 [hep-ph]}
  \BibitemShut {NoStop}%
\bibitem [{\citenamefont {Agostinelli}\ \emph {et~al.}(2003)\citenamefont
  {Agostinelli} \emph {et~al.}}]{AGOSTINELLI2003250}%
  \BibitemOpen
  \bibfield  {author} {\bibinfo {author} {\bibfnamefont {S.}~\bibnamefont
  {Agostinelli}} \emph {et~al.},\ }\href {\doibase
  https://doi.org/10.1016/S0168-9002(03)01368-8} {\bibfield  {journal}
  {\bibinfo  {journal} {Nuclear Instruments and Methods in Physics Research
  Section A: Accelerators, Spectrometers, Detectors and Associated Equipment}\
  }\textbf {\bibinfo {volume} {506}},\ \bibinfo {pages} {250 } (\bibinfo {year}
  {2003})}\BibitemShut {NoStop}%
\bibitem [{\citenamefont {{Yoon}}\ \emph {et~al.}(2011)\citenamefont {{Yoon}},
  \citenamefont {{Ahn}}, \citenamefont {{Allison}}, \citenamefont {{Bagliesi}}
  \emph {et~al.}}]{2011ApJ...728..122Y}%
  \BibitemOpen
  \bibfield  {author} {\bibinfo {author} {\bibfnamefont {Y.~S.}\ \bibnamefont
  {{Yoon}}}, \bibinfo {author} {\bibfnamefont {H.~S.}\ \bibnamefont {{Ahn}}},
  \bibinfo {author} {\bibfnamefont {P.~S.}\ \bibnamefont {{Allison}}}, \bibinfo
  {author} {\bibfnamefont {M.~G.}\ \bibnamefont {{Bagliesi}}},  \emph
  {et~al.},\ }\href {\doibase 10.1088/0004-637X/728/2/122} {\bibfield
  {journal} {\bibinfo  {journal} {Astrophys. J.}\ }\textbf {\bibinfo {volume}
  {728}},\ \bibinfo {eid} {122} (\bibinfo {year} {2011})},\ \Eprint
  {http://arxiv.org/abs/1102.2575} {arXiv:1102.2575 [astro-ph.HE]} \BibitemShut
  {NoStop}%
\bibitem [{\citenamefont {{Adriani}}\ \emph {et~al.}(2011)\citenamefont
  {{Adriani}}, \citenamefont {{Barbarino}}, \citenamefont {{Bazilevskaya}},
  \citenamefont {{Zverev}} \emph {et~al.}}]{2011Sci...332...69A}%
  \BibitemOpen
  \bibfield  {author} {\bibinfo {author} {\bibfnamefont {O.}~\bibnamefont
  {{Adriani}}}, \bibinfo {author} {\bibfnamefont {G.~C.}\ \bibnamefont
  {{Barbarino}}}, \bibinfo {author} {\bibfnamefont {G.~A.}\ \bibnamefont
  {{Bazilevskaya}}}, \bibinfo {author} {\bibfnamefont {V.~G.}\ \bibnamefont
  {{Zverev}}},  \emph {et~al.} (\bibinfo {collaboration} {PAMELA}),\ }\href
  {\doibase 10.1126/science.1199172} {\bibfield  {journal} {\bibinfo  {journal}
  {Science}\ }\textbf {\bibinfo {volume} {332}},\ \bibinfo {pages} {69}
  (\bibinfo {year} {2011})},\ \Eprint {http://arxiv.org/abs/1103.4055}
  {arXiv:1103.4055 [astro-ph.HE]} \BibitemShut {NoStop}%
\bibitem [{\citenamefont {{Panov}}\ \emph {et~al.}(2009)\citenamefont
  {{Panov}}, \citenamefont {{Adams}}, \citenamefont {{Ahn}}, \citenamefont
  {{Bashinzhagyan}}, \citenamefont {{Fazely}} \emph
  {et~al.}}]{2009BRASP..73..564P}%
  \BibitemOpen
  \bibfield  {author} {\bibinfo {author} {\bibfnamefont {A.~D.}\ \bibnamefont
  {{Panov}}}, \bibinfo {author} {\bibfnamefont {J.~H.}\ \bibnamefont
  {{Adams}}}, \bibinfo {author} {\bibfnamefont {H.~S.}\ \bibnamefont {{Ahn}}},
  \bibinfo {author} {\bibfnamefont {G.~L.}\ \bibnamefont {{Bashinzhagyan}}},
  \bibinfo {author} {\bibfnamefont {A.~R.}\ \bibnamefont {{Fazely}}},  \emph
  {et~al.},\ }\href {\doibase 10.3103/S1062873809050098} {\bibfield  {journal}
  {\bibinfo  {journal} {Bulletin of the Russian Academy of Sciences, Physics}\
  }\textbf {\bibinfo {volume} {73}},\ \bibinfo {pages} {564} (\bibinfo {year}
  {2009})},\ \Eprint {http://arxiv.org/abs/1101.3246} {arXiv:1101.3246
  [astro-ph.HE]} \BibitemShut {NoStop}%
\bibitem [{\citenamefont {Aguilar}\ \emph
  {et~al.}(2015{\natexlab{a}})\citenamefont {Aguilar} \emph
  {et~al.}}]{PhysRevLett.114.171103}%
  \BibitemOpen
  \bibfield  {author} {\bibinfo {author} {\bibfnamefont {M.}~\bibnamefont
  {Aguilar}} \emph {et~al.} (\bibinfo {collaboration} {AMS Collaboration}),\
  }\href {\doibase 10.1103/PhysRevLett.114.171103} {\bibfield  {journal}
  {\bibinfo  {journal} {Phys. Rev. Lett.}\ }\textbf {\bibinfo {volume} {114}},\
  \bibinfo {pages} {171103} (\bibinfo {year} {2015}{\natexlab{a}})}\BibitemShut
  {NoStop}%
\bibitem [{\citenamefont {Th{\'e}bault}\ \emph {et~al.}(2015)\citenamefont
  {Th{\'e}bault} \emph {et~al.}}]{Thebault2015}%
  \BibitemOpen
  \bibfield  {author} {\bibinfo {author} {\bibfnamefont {E.}~\bibnamefont
  {Th{\'e}bault}} \emph {et~al.},\ }\href {\doibase 10.1186/s40623-015-0228-9}
  {\bibfield  {journal} {\bibinfo  {journal} {Earth, Planets and Space}\
  }\textbf {\bibinfo {volume} {67}},\ \bibinfo {pages} {79} (\bibinfo {year}
  {2015})}\BibitemShut {NoStop}%
\bibitem [{\citenamefont {{G{\'o}rski}}\ \emph {et~al.}(2005)\citenamefont
  {{G{\'o}rski}}, \citenamefont {{Hivon}}, \citenamefont {{Banday}},
  \citenamefont {{Wandelt}}, \citenamefont {{Hansen}}, \citenamefont
  {{Reinecke}},\ and\ \citenamefont {{Bartelmann}}}]{2005ApJ...622..759G}%
  \BibitemOpen
  \bibfield  {author} {\bibinfo {author} {\bibfnamefont {K.~M.}\ \bibnamefont
  {{G{\'o}rski}}}, \bibinfo {author} {\bibfnamefont {E.}~\bibnamefont
  {{Hivon}}}, \bibinfo {author} {\bibfnamefont {A.~J.}\ \bibnamefont
  {{Banday}}}, \bibinfo {author} {\bibfnamefont {B.~D.}\ \bibnamefont
  {{Wandelt}}}, \bibinfo {author} {\bibfnamefont {F.~K.}\ \bibnamefont
  {{Hansen}}}, \bibinfo {author} {\bibfnamefont {M.}~\bibnamefont
  {{Reinecke}}}, \ and\ \bibinfo {author} {\bibfnamefont {M.}~\bibnamefont
  {{Bartelmann}}},\ }\href {\doibase 10.1086/427976} {\bibfield  {journal}
  {\bibinfo  {journal} {Astrophys. J.}\ }\textbf {\bibinfo {volume} {622}},\
  \bibinfo {pages} {759} (\bibinfo {year} {2005})},\ \Eprint
  {http://arxiv.org/abs/astro-ph/0409513} {astro-ph/0409513} \BibitemShut
  {NoStop}%
\bibitem [{\citenamefont {{Atkins}}\ \emph {et~al.}(2003)\citenamefont
  {{Atkins}} \emph {et~al.}}]{2003ApJ...595..803A}%
  \BibitemOpen
  \bibfield  {author} {\bibinfo {author} {\bibfnamefont {R.}~\bibnamefont
  {{Atkins}}} \emph {et~al.},\ }\href {\doibase 10.1086/377498} {\bibfield
  {journal} {\bibinfo  {journal} {Astrophys. J.}\ }\textbf {\bibinfo {volume}
  {595}},\ \bibinfo {pages} {803} (\bibinfo {year} {2003})},\ \Eprint
  {http://arxiv.org/abs/astro-ph/0305308} {astro-ph/0305308} \BibitemShut
  {NoStop}%
\bibitem [{\citenamefont {{Li}}\ and\ \citenamefont
  {{Ma}}(1983)}]{1983ApJ...272..317L}%
  \BibitemOpen
  \bibfield  {author} {\bibinfo {author} {\bibfnamefont {T.-P.}\ \bibnamefont
  {{Li}}}\ and\ \bibinfo {author} {\bibfnamefont {Y.-Q.}\ \bibnamefont
  {{Ma}}},\ }\href {\doibase 10.1086/161295} {\bibfield  {journal} {\bibinfo
  {journal} {Astrophys. J.}\ }\textbf {\bibinfo {volume} {272}},\ \bibinfo
  {pages} {317} (\bibinfo {year} {1983})}\BibitemShut {NoStop}%
\bibitem [{\citenamefont {{Feldman}}\ and\ \citenamefont
  {{Cousins}}(1998)}]{1998PhRvD..57.3873F}%
  \BibitemOpen
  \bibfield  {author} {\bibinfo {author} {\bibfnamefont {G.~J.}\ \bibnamefont
  {{Feldman}}}\ and\ \bibinfo {author} {\bibfnamefont {R.~D.}\ \bibnamefont
  {{Cousins}}},\ }\href {\doibase 10.1103/PhysRevD.57.3873} {\bibfield
  {journal} {\bibinfo  {journal} {Phys. Rev. D}\ }\textbf {\bibinfo {volume}
  {57}},\ \bibinfo {pages} {3873} (\bibinfo {year} {1998})},\ \Eprint
  {http://arxiv.org/abs/physics/9711021} {physics/9711021} \BibitemShut
  {NoStop}%
\bibitem [{\citenamefont {Donato}\ \emph {et~al.}(2009)\citenamefont {Donato},
  \citenamefont {Maurin}, \citenamefont {Brun}, \citenamefont {Delahaye},\ and\
  \citenamefont {Salati}}]{PhysRevLett.102.071301}%
  \BibitemOpen
  \bibfield  {author} {\bibinfo {author} {\bibfnamefont {F.}~\bibnamefont
  {Donato}}, \bibinfo {author} {\bibfnamefont {D.}~\bibnamefont {Maurin}},
  \bibinfo {author} {\bibfnamefont {P.}~\bibnamefont {Brun}}, \bibinfo {author}
  {\bibfnamefont {T.}~\bibnamefont {Delahaye}}, \ and\ \bibinfo {author}
  {\bibfnamefont {P.}~\bibnamefont {Salati}},\ }\href {\doibase
  10.1103/PhysRevLett.102.071301} {\bibfield  {journal} {\bibinfo  {journal}
  {Phys. Rev. Lett.}\ }\textbf {\bibinfo {volume} {102}},\ \bibinfo {pages}
  {071301} (\bibinfo {year} {2009})}\BibitemShut {NoStop}%
\bibitem [{\citenamefont {Aguilar}\ \emph
  {et~al.}(2015{\natexlab{b}})\citenamefont {Aguilar} \emph
  {et~al.}}]{PhysRevLett.115.211101}%
  \BibitemOpen
  \bibfield  {author} {\bibinfo {author} {\bibfnamefont {M.}~\bibnamefont
  {Aguilar}} \emph {et~al.} (\bibinfo {collaboration} {AMS Collaboration}),\
  }\href {\doibase 10.1103/PhysRevLett.115.211101} {\bibfield  {journal}
  {\bibinfo  {journal} {Phys. Rev. Lett.}\ }\textbf {\bibinfo {volume} {115}},\
  \bibinfo {pages} {211101} (\bibinfo {year} {2015}{\natexlab{b}})}\BibitemShut
  {NoStop}%
\bibitem [{\citenamefont {{Hampel-Arias}}(2017)}]{2017PhDT........59H}%
  \BibitemOpen
  \bibfield  {author} {\bibinfo {author} {\bibfnamefont {Z.}~\bibnamefont
  {{Hampel-Arias}}},\ }\emph {\bibinfo {title} {{Cosmic Ray Observations at the
  TeV Scale with the HAWC Observatory}}},\ \href@noop {} {Ph.D. thesis},\
  \bibinfo  {school} {The University of Wisconsin - Madison} (\bibinfo {year}
  {2017})\BibitemShut {NoStop}%
\end{thebibliography}%
\end{document}